\documentclass[journal ]{new-aiaa}
\usepackage[utf8]{inputenc}
\usepackage{textcomp}

\usepackage{graphicx}
\usepackage{amsmath}
\usepackage[version=4]{mhchem}
\usepackage{siunitx}
\usepackage{longtable,tabularx}
\usepackage{multirow}
\usepackage{amsmath}
\usepackage{caption}
\usepackage{subcaption}
\usepackage{wrapfig}
\usepackage{float}  
\usepackage{ar}
\usepackage{physics}
\usepackage{ulem}

\title{A perspective on transonic buffet over an unswept finite wing}

\author{Magan Singh\footnote{Postdoctoral Fellow, Aerospace Engineering Department; magans@iitk.ac.in}}
\affil{Indian Institute of Technology Kanpur, Kanpur, Uttar Pradesh, 208016, India}
\author{Kartik Venkatraman\footnote{Associate Professor, Aerospace Engineering Department; kartik@iisc.ac.in}}
\affil{Indian Institute of Science, Bengaluru, Karnataka, 560012, India}

\begin{document}

\maketitle

\begin{abstract}
Transonic buffet is a well-known aerodynamic instability of shock/boundary layer interaction in the transonic regime for aircraft. So far, this phenomenon has typically been investigated by modal and correlation analyses. Here, we present a perspective on low-frequency unsteadiness, of the order of $St \sim \mathcal{O}(-2)$, in transonic buffet, using results from temporal evolution of skin friction lines and correlation analysis on the surface of Benchmark Supercritical Wing (BSCW) with an aspect ratio $\AR = 2$. Skin friction lines and critical point theory are well established to describe 3D separated flows based on critical points---nodes, foci, and saddles. Dynamics of these critical points are found in a certain topology in separation regions for different angles of attack, Mach $0.85$, and Reynolds number $4.491\times10^6$. The topology of critical points consists of pairs of contra-rotating unstable foci that contribute to the mechanism of formation and propagation of buffet cells---pockets of shock foot oscillations. The dynamic nature of these critical points changes the pressure distribution on the surface, which is reflected as wave propagation in correlation plots.
\end{abstract}

\section*{Nomenclature}


{\renewcommand\arraystretch{1.0}
\noindent\begin{longtable*}{@{}l @{\quad=\quad} l@{}}
$c$ & Chord length of wing \\
$c'_{p_i}$ & Fluctuating component of $c_p$ at varying location \\
$c'_{p_r}$ & Fluctuating component of $c_p$ at a reference location \\
$e$ & Sum of internal and kinetic energies per unit volume \\
$f$ & Buffet frequency in Hz \\
$\lambda_{bp}$ & Wavelength of buffet cells propagation measured on skin-friction lines plots \\
$\lambda_{pwp}$ & Wavelength of pressure wave propagation computed from cross-correlation plots \\
${\tilde{L}_{R}}$ & Relative characteristic length \\
$Pr$ & Prandtl number \\
$p$ & Pressure \\
$q_{x_i}$ & Heat flux in index notation \\
$Re_c$ & Reynolds number based on chord \\
$Re_{\tilde{L}_{R}}$ & Reynolds number based on ${\tilde{L}_{R}}$ \\
$R_{c'_{p_r}c'_{p_i}}$ & Normalized cross-correlation between $c'_{p_r}$ and $c'_{p_i}$\\
$s$ &  Span length of wing \\
$St$ & Strouhal number ($f c/V_{\infty}$) \\
$T$ & Temperature \\
$t$ & Time \\
$u,v,w$ & Mean component of velocity vector $\mathbf{V}$ in Cartesian coordinates \\
$u',v',w'$ & Fluctuating component of velocity vector $\mathbf{V}$ in Cartesian coordinates \\
$U_{\infty}$ & Free-stream speed \\
$V_p$ & Pressure wave propagation velocity \\
\multicolumn{2}{@{}l}{Greek alphabets}\\
$\gamma$ &  Ratio of specific heats \\
$\mu$ & Fluid dynamic viscosity \\
$\rho$ & Fluid density \\
$\tau$ & Time lag between two signals in cross-correlation \\
$\tau_{x_ix_j}$ & Shear stress in index notation \\
\multicolumn{2}{@{}l}{Subscripts}\\
$\infty$ & Represent free-stream flow \\
$l$ & Represent laminar flow \\
$t$	& Represent turbulent flow
\end{longtable*}}

\section{Introduction} 
\label{sec:intro}

\lettrine{T}{ransonic} shock buffet---shock/boundary layer interaction--- has been investigated for more than two decades, both, through analysis of data from wind tunnel experiments, and high-fidelity numerical simulations. Phenomenologically, transonic buffet can be categorized into those that occur in laminar to turbulent transitional flows, and those that occur in fully turbulent flows. Within each of these sub-categories, one could further classify transonic buffet in terms of its occurrence in either 2D or 3D flows. On the other hand, studies on transonic buffet in the literature can be compartmentalized broadly into two groups---modal and correlation analyses. The tools of choice in modal analysis are proper orthogonal decomposition (POD), spectral proper orthogonal decomposition (SPOD), dynamic mode decomposition (DMD), resolvent analysis, and global stability analysis. POD, SPOD, and DMD have been used to identify the dominant spatial patterns and their frequencies in the buffet flow \cite{dandois2018large,poplingher2019modal,masini2020analysis}. Global stability analysis uses a linearized form of RANS equations for the unsteady perturbation about, either the mean flow, or steady flow \cite{crouch2007predicting,timme2020,sansica2023}. The interest here is to determine the dominant unstable eigenvalues that cause shock buffet, and the spatial contours of the corresponding fluid mode. The second group includes studies that use correlation analysis in the time and frequency domain in order to determine regions of coherence in space and time that denote wave propagation pathways in the fluid \cite{deck2005,jacquin2009,hartmann2013interaction}. \citet{lee1990transonic} used the latter analysis to confirm a feedback loop mechanism for self-sustained shock oscillations on a 2D configuration. This mechanism is based on acoustic waves traveling within and outside the boundary layer between the shock and trailing edge. However, \citet{paladini2019various} and \citet{moise2024connecting} refuted this mechanism based on the redundancy of shock in buffet instability. For more details on transonic buffet, there are excellent literature reviews \citep{giannelis2017,gao2020} that cover the mechanisms that cause transonic buffet, the stability perspective on transonic buffet, and the numerous experimental and numerical simulations on transonic buffet and buffeting. In the present study, transonic buffet on an unswept wing is investigated from a very physical perspective. A critical point theory \cite{delery2001annual} is used to discern the critical points---saddles, nodes, and foci---in the flow separation region, and cross-correlation analysis is used for wave propagation in specified regions.

In literature, experimental and numerical studies were conducted in order to understand the flow physics and characteristics of 3D buffet over wings. 3D buffet on swept-wings differs from 2D buffet in two main aspects: first, 3D buffet consists of a broadband frequency content in the range of Strouhal numbers $0.2$-$0.6$ \citep{dandois2016experimental,dandois2019}, which is typically one order higher than the single dominant frequencies of 2D buffet---$0.05$-$0.08$ \citep{jacquin2009}; and second, 3D buffet amplitude, or chord-wise shock excursion, varies along the wing span and is usually much smaller than the 2D buffet amplitude. In 3D buffet, \citet{iovnovich2015} used URANS simulations for capturing buffet on infinite straight and swept wings and introduced the term `buffet cells,' for outboard propagating pockets of chordwise shock oscillations along the span. They found buffet behavior similar to the 2D phenomenon at zero or small sweep angles and also found well-predicted Strouhal numbers of the 3D buffet as the sweep increases. Using scale-resolving DES and URANS, \citet{sartor2017delayed} captured the span-wise outboard propagation of buffet cells in the same range of Strouhal numbers for a swept wing as mentioned above. Both, DES and URANS, agree well in characterizing the 3D buffet. Using 2.5D URANS, \citet{plante2020similarities} linked buffet cells on infinite straight and swept wings with stall cells appearing in a subsonic regime. They found the superposition of 3D buffet (buffet cells) with 2D buffet (shock oscillations) phenomenon, as both phenomena' frequencies are present in the flow for all the mentioned sweep angles. From these studies, it is very clear that 3D buffet is quite a distinct phenomenon from 2D buffet, and also URANS is a reasonable model for the prediction of the 3D buffet on wings. Further, spanwise inboard and outboard convection of buffet cells have been found in coexistence on a swept wing by \citet{masini2020analysis} from experimental data. Inboard propagating perturbations of low frequency are highlighted in Strouhal number range $0.05$-$0.15$ comparable to 2D buffet, and outboard propagating perturbations of high-frequency range $0.2$-$0.5$. So far, in our discussion, the literature is focused on 3D buffet with respect to 2D buffet. The literature on 2D buffet includes most of the studies either supporting \cite{deck2005,xiao2006numerical,jacquin2009,hartmann2013interaction} \citeauthor{lee1990transonic}'s mechanism or opposing \cite{paladini2019various,moise2024connecting} it using different methods. In the end, literature is very short in the inboard propagation of pressure perturbations or buffet cells in 3D buffet flows. The present study finds insights into the mechanism of inboard propagation of buffet cells, and the frequencies of these perturbations are in a similar range of Strouhal numbers as found by \citet{masini2020analysis}.

The present study is arranged as follows. Test case details of the finite span wing in \autoref{sec:method} are first described, followed by governing equations, numerical methods, computational mesh and boundary conditions, critical point theory, and cross-correlation analysis in \autoref{sec:mt} of methods and tools. Thereafter follows the unsteady results in \autoref{sec:rd}, wherein we first present the time and frequency domain analyses in \autoref{subsec:td} and \autoref{subsec:fa}, respectively. \autoref{sec:ch4_shck_sfl_prs} follows a detailed discussion of the nature of pressure wave propagation, chordwise as well as spanwise, and relates it to temporal pressure distribution and topology of skin friction lines. Then \autoref{sec:discuss} provides a detailed discussion about the results, followed by the details of insights into physical mechanisms for self-sustained propagation of buffet cells based on the presence of critical points, in \autoref{ssec:shockBuffMech}. Finally, \autoref{sec:conclu} concludes the whole study.

\section{Test case}
\label{sec:method}

In this study, we numerically simulate the transonic flow past Benchmark Supercritical Wing (BSCW) at different angles of attack (AoAs), namely, $-1^\circ$, $0^\circ$, $1^\circ$, $3^\circ$, $5^\circ$, and $7^\circ$, at free-stream conditions Mach $0.85$ and Reynolds number $4.491\times 10^6$. The same configuration \citep[Figure 1]{heeg2015} and flow conditions \citep[Table 3 and 4]{heeg2015} were used in the Aeroelastic Prediction Workshop-2 \citep{aepw2} (AePW-2) for benchmark studies. In AePW-2, these free-stream conditions, named "Optional Test Case-3a," were identified with strong shock waves and flow separation \citep{chwalowski2016,heeg2013b}. BSCW configuration is a uniform wing with a rectangular planform area, having an aspect ratio of $2$, and whose cross-section is based on a second-generation NASA supercritical airfoil---NASA SC(2)-0414---with a design lift coefficient of $0.4$ and a maximum thickness to chord ratio of $14\%$ \citep{dansberry1992dynamic}. For this test case, the experimental results obtained in a transonic dynamic tunnel are available from the AePW-2 workshop \citep{aepw2}.

\section{Methods and tools}
\label{sec:mt}
The aerodynamic simulations were carried out in CFL3D v6.7 \cite{cfl3d2017,cfl3dmanual,bartels2006cfl3d}, an open-source CFD code. Below, we present the governing equations, numerical methods used in the present CFD simulations, details on the domain discretization and boundary conditions, critical point theory, and cross-correlation analysis.

\subsection{Governing equations}
\label{subsec:goveqn}

Reynolds-averaged Navier-Stokes (RANS) equations are solved for steady and unsteady fully turbulent compressible flows. These equations are non-dimensionalized using free-stream parameters mentioned in \cite[Chapter 4]{cfl3dmanual}. The non-dimensionalized RANS equations \cite[Appendix F]{cfl3dmanual} are written in vector form as follows.
\begin{equation}
\label{eq:rans}
\frac{\partial \mathbf{Q}}{\partial t} + \frac{\partial \mathbf{f}}{\partial x} + \frac{\partial \mathbf{g}}{\partial y} + \frac{\partial \mathbf{h}}{\partial z}  = 0,
\end{equation} 

Where, $\mathbf{Q}$ is a vector of conserved variables---density $\rho$, momentum $\rho u_i$, and total energy $e$ per unit volume. $\mathbf{f}$, $\mathbf{g}$, and $\mathbf{h}$ vectors contain the inviscid, viscous, and heat transfer flux terms in the $x,y$, and $z$ directions, respectively. \autoref{eq:rans} are not closed because of unknown terms. We need to define pressure $p$, shear stress $\tau_{x_i x_j}$, and heat flux $q_{x_i}$ in terms of the primitive variables as
\begin{align}
\label{eq:PrsStrsHeatFlux}
\begin{split}
& p = \left( \gamma -1 \right) \left[ e - \frac{1}{2} \rho u_i u_{i} \right], \\
& \tau_{x_i x_j}= \frac{M_{\infty}}{Re_{\tilde{L}_{R}}} (\mu_l+\mu_t) \left[ \left( \frac{\partial{u_{j}}}{\partial{x_{i}}} + \frac{\partial{u_{i}}}{\partial{x_{j}}} \right) - \frac{2}{3} \frac{\partial{u_{k}}}{\partial{x_{k}}} \delta_{i j} \right], \\
& q_{x_i} = -  \frac{M_{\infty}}{Re_{\tilde{L}_{R}} \left( \gamma - 1 \right)} \left[ \frac{\mu_l}{Pr_l} + \frac{\mu_t}{Pr_t} \right] \frac{\partial{T}}{\partial{x_{i}}}.
\end{split}
\end{align}
In the above, $Re_{\tilde{L}_{R}}=\tilde{\rho}_{\infty}\tilde{U}_{\infty}\tilde{L}_{R}/\tilde{\mu}_{\infty}$. Further, $\tilde{L}_{R}=\tilde{L}/L_{ref}$, where $\tilde{L}$ is a characteristic length; in this study, it is BSCW's chord length. $L_{ref}$ is the corresponding characteristic length in the grid. If a grid is generated in a different length unit other than the wing model, then both, $\tilde{L}$ and $L_{ref}$, must be different; otherwise, they are the same. The laminar dynamic viscosity $\mu_l$ and turbulent eddy viscosity $\mu_t$ are derived using Sutherland's law and the Spalart-Allmaras (SA) turbulence model \cite[Appendix H.2]{cfl3dmanual}, respectively. Finally, the \autoref{eq:rans} and \autoref{eq:PrsStrsHeatFlux} are closed by determining temperature $T$ from the equation of state $ T = \gamma p/\rho$. Further, non-tilde and tilde quantities denote non-dimensional and dimensional quantities, respectively.

\subsection{Numerical methods}
\label{subsec:nm}

For spatial discretization of RANS, a semi-discrete finite-volume scheme \cite[Appendix A]{cfl3dmanual} is used on a structured grid. The convective and pressure terms are differenced using Roe's second-order upwind flux-difference-splitting technique \cite{roe1981approximate}. A smooth limiter \cite[Appendix C]{cfl3dmanual} tuned to a third-order upwind scheme ensures the monotone interpolation of conserved variables $Q$ at interfaces. Shear stress and heat transfer terms are discretized using a second-order central difference scheme. The equations are temporally discretized using Euler's second-order backward implicit scheme. The time advancement is done in the dual time stepping technique. In that, a sufficient number of sub-iterations are carried out for every time step to reduce the residue---typically below $4$ orders of magnitude.

\subsection{Computational mesh and boundary conditions}
\label{subsec:cm}

\begin{figure}[!h]
\centering
\begin{subfigure}{0.49\textwidth}
\centering
\includegraphics[width=0.54\textwidth,height=0.73\textwidth]{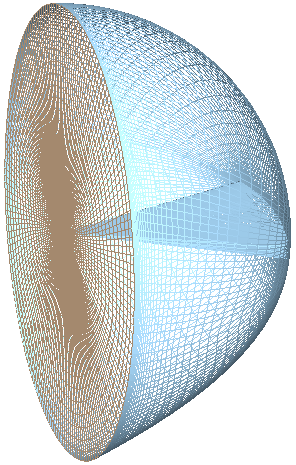}
\caption{Grid topology}  
\label{subfig:grid}
\end{subfigure}
\hfill
\begin{subfigure}{0.49\textwidth}
\centering
\includegraphics[width=0.83\textwidth,height=0.7\textwidth]{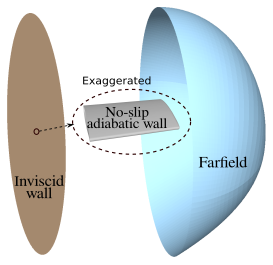}
\caption{Boundary conditions}  
\label{subfig:bcs}
\end{subfigure}
\caption{Grid topology and boundary conditions} 
\label{fig:gridANDbcs}
\end{figure}

Simulations are performed on the structured grids available from AePW-2 \citep{aepw2}. The grids are in three different sizes, namely, a coarse grid with  $1.4$ million nodes, medium grid with $4.8$, and fine grid with $16.9$. Both surfaces of the wing are discretized by $87\times69$, $129\times105$, and $193\times161$ in chord and span for coarse, medium, and fine grids, respectively. Every grid is composed of $5$ blocks. The whole grid is in the form of a hemispherical topology, as shown in \autoref{subfig:grid}. The hemispherical boundary or far-field is at $100c$ from the wing's surface. The $Y^+$ values for coarse, medium, and fine grids are $1$, $2/3$, and $4/9$, respectively. They correspond to first cell heights of $2.39\times10^{-6}$~m, $1.60\times10^{-6}$~m, and $1.06\times10^{-6}$~m, respectively. Further details about the grids and grid convergence study can be found in \citet{singh2022transonic}. The fine grid is chosen for all the simulations because of the higher mesh resolution in the shock excursion and flow separation regions, even though the medium grid also captured inherent unsteadiness and almost converged with the fine grid in steady flow simulations \citep[Figure 8]{singh2022transonic}.

Boundary conditions are set on the grid boundaries as shown in \autoref{subfig:bcs}. Inviscid wall, no-slip adiabatic wall, and far-field boundary conditions are set at the plane, wing surface, and hemispherical boundaries, respectively. Inviscid wall is used to model the splitter plate \cite[Figure 1a]{heeg2015} on which BSCW is placed in transonic dynamic tunnel.

\subsection{Critical point theory}
\label{subsec:cpt}

\begin{figure}[h!]
\centering
\includegraphics[width=\linewidth]{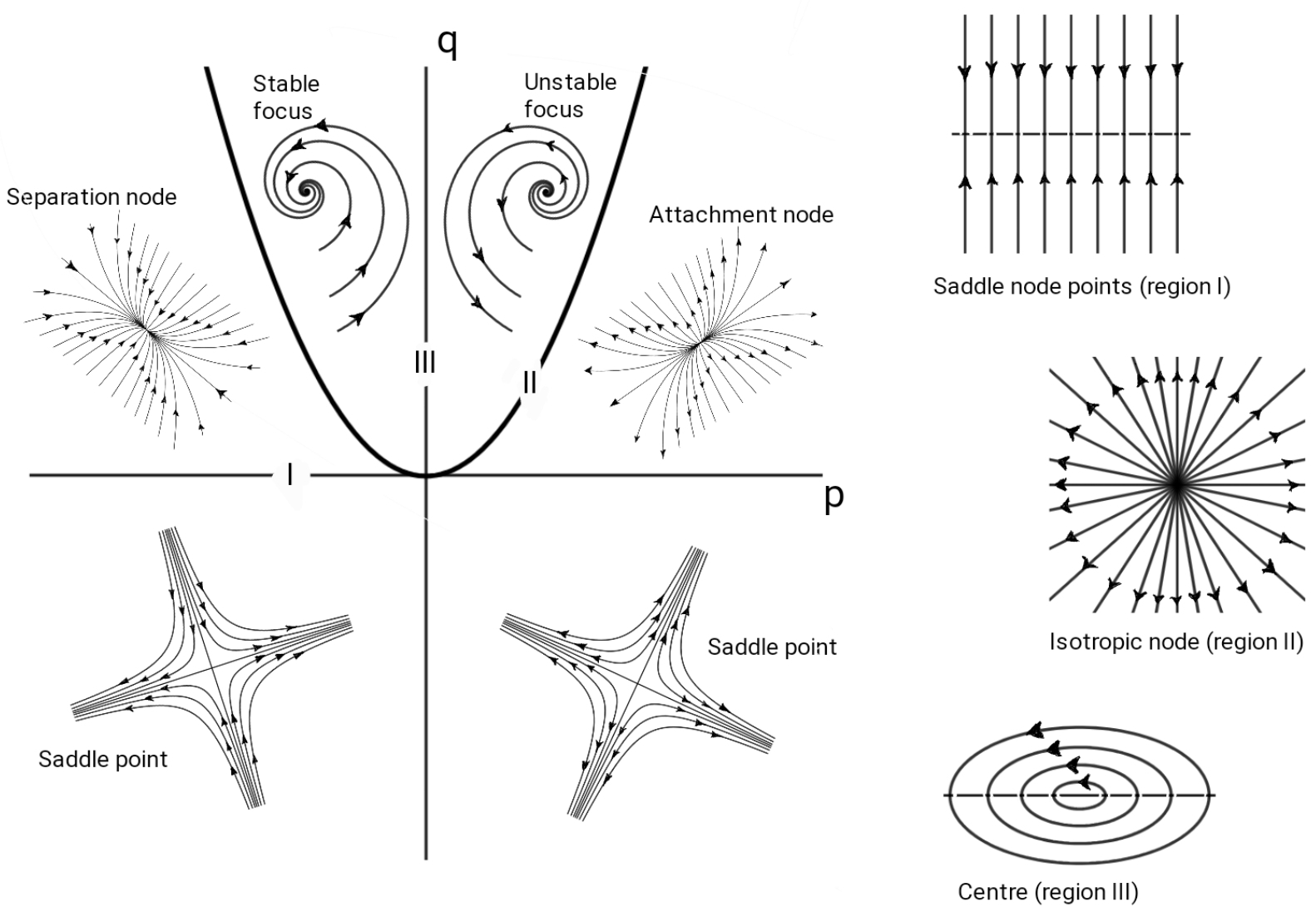}
\caption{Critical points (adapted from \citet[Figure 2 and 3]{delery1992})} 
\label{fig:crit_points}
\end{figure}

The visualization of skin friction lines and their use in describing and interpreting the nature of flow separation and re-attachment on a solid wall has been well discussed by \citet{delery2001annual}, and \citet{delery2013}, among many others. This interpretation is based on critical point theory that describes the flow topology about critical points given in \autoref{fig:crit_points}, such as nodes, foci, and saddle points. Critical point theory is devised from the mathematical work of Poincare (1892) on the equilibrium points of differential equations of dynamical systems. In words of \citet{delery2001annual}, "The present theory of three-dimensional separation is not a predictive theory in that it does not predict separation on an obstacle in given conditions. \ldots It is a descriptive theory reasoning on the properties of a given vector field (skin friction or velocity) provided by experimentation or calculation." For a given skin-friction vector field on the surface, the concepts from critical point theory can be applied to discern the unstable foci in a certain topology of critical points, causing the shock buffet. In the following paragraph, we briefly review the mathematical description given in the theory for critical points.

Skin friction lines are curves whose tangent at any given point denotes the shear stress vector on the wall surface. Flow streamlines above the surface coincide with the skin friction lines in the limit when the wall-normal distance tends to zero. This is because skin-friction vector is proportional to the normal derivative of the velocity vector. The skin friction lines in the field of skin friction vector $\tau=\tau_x (x,z)~i + \tau_z (x,z)~k$ on the surface are defined by the following differential system \citep[Section 1.2]{delery2013}.

\begin{equation} \label{eq:deOFsfls}
    \frac{dx}{\tau_x (x,z)}=\frac{dz}{\tau_z (x,z)}.
\end{equation}

Two or more skin-friction lines can not intersect at a point, similar to streamlines. Points, where skin-friction vector $\tau$ vanishes, are called critical points $P_0(x,z)$ of the system. The differential system turns into the following eigenvalue problem at points $P_0$ as explained in \cite[Section 1.3.1]{delery2013}.

\begin{equation}
\label{eq:critpnt_shrstrs_chardir}
\begin{bmatrix}
\pdv{\tau_x}{x} - S & \pdv{\tau_z}{x} \\
\pdv{\tau_x}{z} & \pdv{\tau_z}{z} - S
\end{bmatrix}
\begin{bmatrix}
\lambda \\ \mu
\end{bmatrix}
=
\begin{bmatrix}
0 \\ 0
\end{bmatrix}
\end{equation}

In \autoref{eq:critpnt_shrstrs_chardir}, $S$ denotes the characteristic value, or eigenvalue, and $[\lambda,\mu]^T$ the components of the characteristic direction, or eigenvector. Non-trivial values of the characteristic direction are determined by setting the determinant of the $2\times2$ matrix in \autoref{eq:critpnt_shrstrs_chardir} equal to zero. There are two characteristic directions, or eigenvectors, corresponding to the eigenvalues $S_1$ and $S_2$. The characteristic equation that determines the eigenvalues is of the form $S^2 - pS + q = 0$. $p$ and $q$ are related to the eigenvalues as $S_1 + S_2 = p$ and $S_1 S_2 = q$. The parameters $p$ and $q$ constitute a plane of classifying the different regions of critical points as shown in \autoref{fig:crit_points}. Each critical point is explained in detail with its mathematical description in \cite[Section 1.3.2]{delery2013}. For different sets of $p$ and $q$ values, critical points lie in different quadrants of the $p-q$ plane. Among all critical points, unstable foci in a certain topology play an important role in the self-sustained propagation of buffet cells, as discussed in the present study. The unstable focus lies in a region between the parabola $p^2=4q$ (region II) and $q=0$ (region III) in the first quadrant and stable in the second quadrant.

\citet[Section 3.2]{delery2013} describe a relation between the total number of nodes $N$, foci $F$, and saddles $S$ on the surface of a wing mounted on a wall in a wind tunnel given by 
\begin{equation}
\label{eq:saddle_nodesfoci_rel}
    N + F = S.
\end{equation}
Therefore, one could have a vector field with no saddles or nodes and foci at all or have an equal number of saddles and the sum of nodes and foci. This relation confirms the invariance of structural stability of the topology of skin friction lines upon a small change in a flow parameter, such as the angle of attack, Mach number, or Reynolds number \citep{tobak1982}. The same relation is valid for every instance of the topology of skin-friction lines at every AoA from the present URANS simulations.

\subsection{Cross-correlation analysis}
\label{subsec:cra}

The cross-correlations of fluctuating component $c'_p$ of the coefficient of pressure $c_p$ are computed to analyze wave propagations on the surfaces. For discrete points in space and time, the two-point cross-correlation is expressed as

\begin{equation}
\label{eq:cca2}
\hat{R}_{s_1s_2}(X,\tau) = \sum_i s_1(X_r,t_i)~s_2(X,t_i+\tau)\ \delta t.
\end{equation}

Where $\tau$ is a time lag between signals $s_1$ and $s_2$. $\delta t$ is the time interval between two consecutive values of a signal. $X_r$ is a reference position vector, and $X$ is a varying position vector in space. In the present study, correlation is computed along the chordwise and spanwise lengths on the suction and pressure surfaces as

\begin{equation}
R_{c'_{p_r}c'_{p_i}}(X,\tau) =  \frac{1}{N\delta t}\hat{R}_{c'_{p_r}c'_{p_i}}(X,\tau).
\end{equation}

Where $c'_{p_r}$ and $c'_{p_i}$ are the fluctuating components of $c_p$ at a reference and any other varying location along chord or span, respectively. $N$ is the total number of samples of $c_p$ time history. For brevity, we will represent $R_{c'_{p_r}c'_{p_i}}$ as $R_{ri}$ in further text and figures. In figures, $R_{ri}(X,\tau)$ contours are plotted with $\tau$ as $x$-axis and chord or span length $X$ as $y$-axis.

\section{Unsteady Results}
\label{sec:rd}

All steady and unsteady RANS simulations are performed on the fine grid of $16.9$ million nodes. Simulations are performed at a time step $dt = 7.535 \times 10^{-7}~\text{sec}$. The grid and time step are selected after convergence studies with the same configuration and flow conditions in our earlier study \cite{singh2022transonic}. URANS simulations are restarted from last iteration of the steady flow simulations for all AoAs; density residue drops below the order of $4$ at every time step in $10$ sub-iterations. The numerical results are compared with the experimental results available from AePW-2 \cite{aepw2}. The subsequent subsections analyze unsteady results using time domain, frequency domain, and cross-correlation analyses.

\subsection{Time domain analysis} 
\label{subsec:td}

\begin{figure}[!h]
\centering
\begin{subfigure}{0.49\textwidth}
\centering
\includegraphics[width=\linewidth]{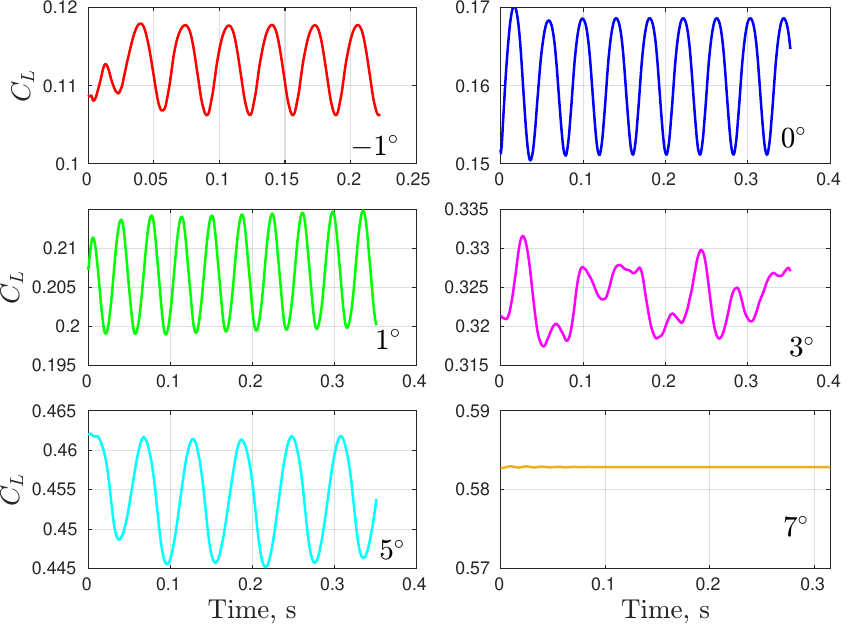}
\caption{$C_L$ versus time}  
\label{subfig:clOscill}
\end{subfigure}
\begin{subfigure}{0.49\textwidth}
\centering
\includegraphics[width=\linewidth]{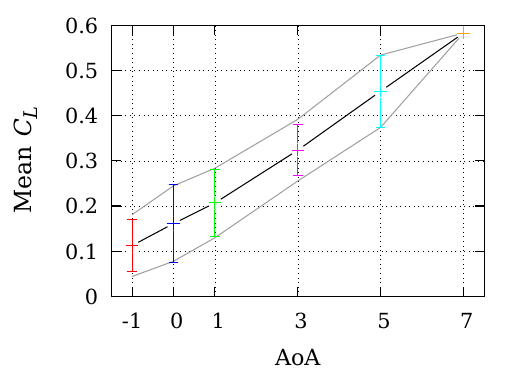}
\caption{Mean $C_L$ versus AoA}  
\label{subfig:clVsAlpha}
\end{subfigure}
\caption{Unsteady simulation results} 
\label{fig:unStdySim}
\end{figure}

The simulations at Mach $0.85$ and Reynolds number $4.491\times10^6$ predict oscillations in the wing's coefficient of lift $C_L$ at all AoAs except $7^\circ$ shown in \autoref{subfig:clOscill}. For $7^\circ$, unsteady simulation converges to a $C_L$ value of $0.583$ and predicts a steady flow. Therefore, buffet offset---an AoA at which buffet ceases---occurs between $5^\circ$ and $7^\circ$.

In \autoref{subfig:clOscill}, $C_L$ oscillations are almost sinusoidal for $-1^\circ$, $0^\circ$, $1^\circ$, and $5^\circ$ AoAs. But for $3^\circ$, oscillations are aperiodic in nature. These oscillations at every AoA are due to the fully developed shock buffet on BSCW. Propagations of buffet cells are observed on the suction surface (SS) and/or pressure surface (PS) in submitted respective supplementary videos for these AoAs. There are a total of 10 animation videos, two for each AoA. For example, Supplementary Videos 1A and 1B are for the suction and pressure surfaces, respectively, for $-1^\circ$; similarly, 2A and 2B are for $0^\circ$ AoA and so on. \autoref{subfig:clVsAlpha} shows the mean $C_L$ variation with respect to AoA. The mean $C_L$ varies linearly with AoA from $-1^\circ$ to $7^\circ$. This almost linear relation confirms that the shock buffet is a pre-stall aerodynamic instability. Peak to peak oscillation amplitudes, $5$ times of their original values, are shown in exaggeration about mean $C_L$ values. The original values are $0.011382$, $0.017377$, $0.014759$, $0.011297$, and $0.016011$ for $-1^\circ$, $0^\circ$, $1^\circ$, $3^\circ$, $5^\circ$ AoAs. This variation of the amplitudes is closely related to the buffet amplitudes (shock foot excursion length) on suction and pressure surfaces at $10\%$ and $60\%$ spans in \autoref{subfig:baVSaoa}, discussed later in the paragraph. Notice that typically for $-1^\circ$, $0^\circ$, and $1^\circ$ AoAs buffet amplitude dominates from PS, for $5^\circ$ it dominates from SS, and for $3^\circ$ it dominates from both surfaces but lower than other AoAs.

\begin{figure}[!h]
\centering
\begin{subfigure}[b]{0.49\textwidth}
\centering
\includegraphics[width=\linewidth]{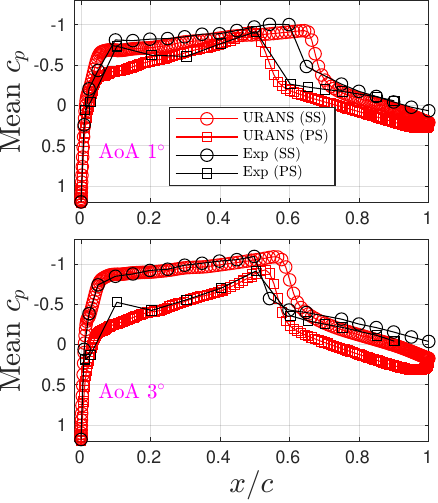}
\caption{}
\label{subfig:meancp}
\end{subfigure}
\hfill
\begin{subfigure}[b]{0.49\textwidth}
\centering
\includegraphics[width=\linewidth]{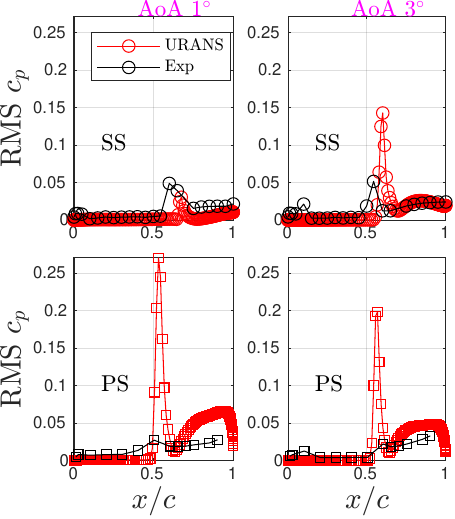}
\caption{}  
\label{subfig:rmscp}
\end{subfigure}
\caption{Comparison between URANS and experiment results} 
\label{fig:Validation}
\end{figure}

\begin{figure}[!h]
\centering
\begin{subfigure}[b]{0.49\textwidth}
\centering
\includegraphics[width=\linewidth]{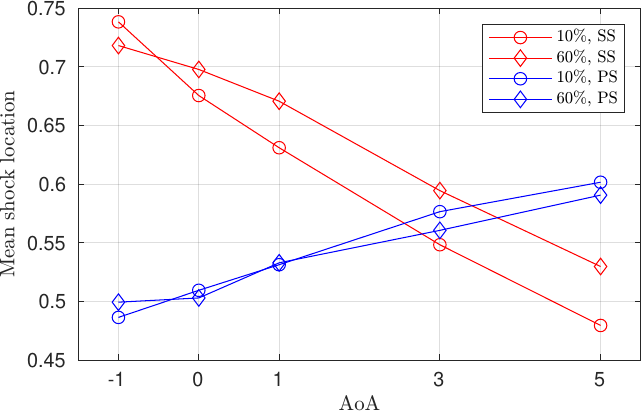}
\caption{Mean shock location}  
\label{subfig:mslVSaoa}
\end{subfigure}
\hfill
\begin{subfigure}[b]{0.49\textwidth}
\centering
\includegraphics[width=\linewidth]{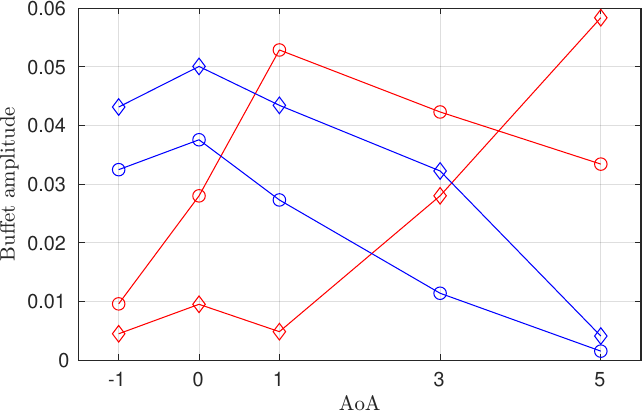}
\caption{Buffet amplitude}  
\label{subfig:baVSaoa}
\end{subfigure}
\caption{Buffet characteristics with respect to AoA} 
\label{fig:bufftCharac}
\end{figure}

For validation, we compared our numerical mean and RMS $c_p$ distributions at $60\%$ span with the experiment. The comparison is done for all buffet-developed AoAs. But here, it is shown for $1^\circ$ and $3^\circ$ in \autoref{fig:Validation}. From both, mean and RMS $c_p$, URANS with the SA turbulence model shows a reasonable agreement and similar trend except for deviation in shock and separation region. The deviation is larger for higher AoAs $3^\circ$ and $5^\circ$ due to strong flow separation than lower ones $-1^\circ$, $0^\circ$, and $1^\circ$. For all AoAs, simulations typically capture the shock downstream relative to the experiment, especially on the suction surface, and overestimate pressure fluctuations in the shock and separation regions. However, from RMS $c_p$ plots, a trend of increase and decrease of pressure fluctuations in the shock and separation regions with the increase of AoA is well captured on the suction and pressure surfaces, respectively.

Mean shock locations and buffet amplitudes, normalized by the chord length $c$, are given as a function of AoA in \autoref{fig:bufftCharac}. They are computed at $10$ and $60\%$ spans. Mean shock location with increasing AoA monotonically moves toward the leading edge on SS and toward the trailing edge on PS. Hence, the separation region increases on SS and decreases on PS, as seen in mean skin-friction lines in \autoref{sec:ch4_shck_sfl_prs}. Buffet amplitude at $60\%$ span on SS remains almost constant till $1^\circ$ AoA and subsequently increases. However, at $10\%$ span, the buffet amplitude increases till $1^\circ$ and then decreases. On PS, the amplitude first increases from $-1^\circ$ to $0^\circ$ and then decreases for both these span stations. Overall, the amplitudes on both surfaces typically vary between $0.5\%$ to $5.5\%$ of the chord on both span stations. These buffet amplitudes are much lower than the buffet amplitude in 2D buffet flow on the airfoil section of BSCW in \citet{arif2023two}.

\subsection{Frequency domain analysis}
\label{subsec:fa}

\begin{figure}[!ht]
\centering
\begin{subfigure}[b]{0.48\textwidth}
\includegraphics[width=\linewidth]{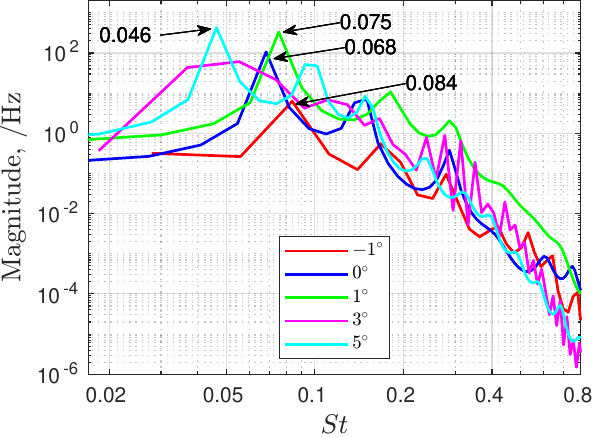}
\caption{}  
\label{subfig:psdsAllAoAs}
\end{subfigure}
\begin{subfigure}[b]{0.48\textwidth}
\includegraphics[width=\linewidth]{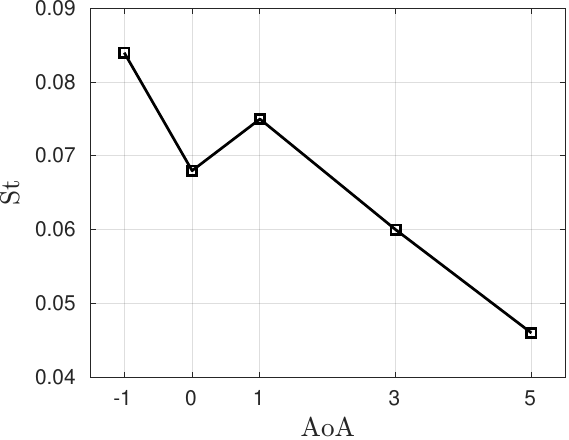}
\caption{}
\label{wrapfig:BufftFrq}
\end{subfigure}
\caption{Frequency domain analysis: (a) PSDs at mean shock locations on suction side at $10\%$ span for each AoA; (b) Srouhal number $St$ variation with increase of AoA}
\label{fig:psd3d}
\end{figure}

The power spectral density (PSD) of pressure fluctuations for buffet-developed AoAs are computed to analyze the frequency and energy content in the unsteady flow. PSDs are determined at the mean shock and $90\%$ chord locations on both surfaces at $10\%$ and $60\%$ spans. PSDs at both span locations provide the same dominant peaks with different energy contents. For all AoAs, PSDs at mean shock locations on the suction side of $10\%$ span are included in \autoref{fig:psd3d}. We use an autoregressive (AR) PSD estimator using the Burg method \citep{burg1967maximum}. The method has been used in studies of transonic buffet---\cite{deck2005,masini2020analysis} for short length signals. In the present study, PSDs are computed with $450$ samples of a signal length $0.10$~sec for all AoAs. This length comprises almost $4$ cycles of $C_L$ for every AoA. PSDs are computed with a single window and an order of $300$ for $-1^\circ$ and $3^\circ$, and an order of $50$ for other AoAs. In plots, the frequency is expressed in terms of Strouhal number defined as $St=fc/U_{\infty}$, where $f$ is the frequency in Hz, chord length $c=16$ inches, and the free stream speed $U_{\infty}=468.73$ ft/s.

PSDs predict a dominant peak at $St=$ $0.084$, $0.068$, $0.075$, and $0.046$ for $-1^\circ$, $0^\circ$, $1^\circ$, and $5^\circ$ AoAs, respectively. For $3^\circ$, PSD predicts a band of $St$ $0.04-0.08$. A dominant frequency relates to sinusoidal oscillations of $C_L$ for $-1^\circ$, $0^\circ$, $1^\circ$, and $5^\circ$ AoAs and a band to aperiodic oscillations for $3^\circ$. Strouhal number $St$ versus AoA in \autoref{wrapfig:BufftFrq} shows a typical decrease of $St$ with increase of AoA, which is contrary to the increase of $St$ with increase of AoA in numerical 2D buffet studies by \citet{crouch2009} and \citet{arif2023two}.


\subsection{Cross-correlations, skin friction lines, and temporal $c_p$ distributions}
\label{sec:ch4_shck_sfl_prs}

The correlations are computed to examine the velocity $V_p$ of pressure wave propagations along streamwise and spanwise lengths on the suction and pressure surfaces. Streamwise correlations are computed at $10\%$ and $60\%$ spans on both surfaces. Spanwise correlations are computed along the spanwise lines through the shock excursion region from the root to $90\%$ span on both surfaces. These spanwise straight lines are taken at chord locations $x/c$ (mentioned in spanwise correlation plots) of higher pressure fluctuations along the span. The streamwise correlations at $10\%$ span are only presented here for all AoAs, as this region exhibits strong separation and richness of critical points. Whereas, the $60\%$ span lacks the same. The reference locations $x_r/c$ for streamwise correlations and $y_r/s$ for spanwise correlations are mentioned in their respective figures.

The wave propagation velocity $V_p$ is calculated as the slope of a trace of local maxima of correlation function $R_{ri}$ in the same phase at different streamwise or spanwise locations. A positive slope indicates wave propagation along the positive spatial direction. That means towards the trailing edge (TE) in streamwise correlation and towards the wing's tip in spanwise correlation, contrary to the negative slope. The propagation velocity or phase velocity value is computed in m/s (SI units) and is shown on the correlation plots. From these plots, two types of wave propagation are identified: hydrodynamic and acoustic. Hydrodynamic is carried at speed (in m/s) of order $\mathcal{O}(1)$ by the self-induced motion of unstable foci and acoustic at speed of order $\mathcal{O}(2)$ or more. The present research focuses on the hydrodynamic waves as they are simultaneously associated with the self-induced motion of foci and propagation of buffet cells. The distance between two consecutive slopes on the correlation plot denotes the propagating wave's time period. The inverse of this time period gives the frequency $f$ in Hz, which is annotated on correlation plots in terms of Strouhal number $St$. $St$ from correlation plots for an AoA matches very well with $St$ from respective PSD in \autoref{subsec:fa}. Once the phase speed $|V_p|$ and frequency $f$ are determined; wavelength $\lambda_{pwp}$ of pressure waves propagation in streamwise and spanwise directions are easily determined from $\lambda_{pwp} = |V_p|/f$ or $\lambda_{pwp}/c = (|V_p|/U_{\infty})/St$.

Correlation plots relate to temporal $c_p$ distribution and skin-friction line plots. Hence, they are arranged together with number markings to show the change in one reflected in the other two at marking locations. Here, $c_p$ distribution and streamwise correlation plots are included only for $-1^\circ$ AoA. Typically, these plots provide similar inferences for all AoAs. In another sense, spanwise correlations are of higher magnitude and show stronger propagation, most of them of hydrodynamic nature. The relation between these plots becomes very clear in the supplementary videos. Where we can notice a synchronization between the instantaneous change of $c_p$ contours, propagation of buffet cells, and self-induced motion of unstable foci (refer to \autoref{ssec:shockBuffMech} for discussion of a typical topology consisting of foci) in a respective supplementary video. For every AoA, wave propagations are separately analyzed from correlation plots and their relation to critical points.

\begin{figure}[!h]
\centering
\begin{subfigure}[b]{0.49\textwidth}
\centering
\includegraphics[width=1\textwidth,keepaspectratio]{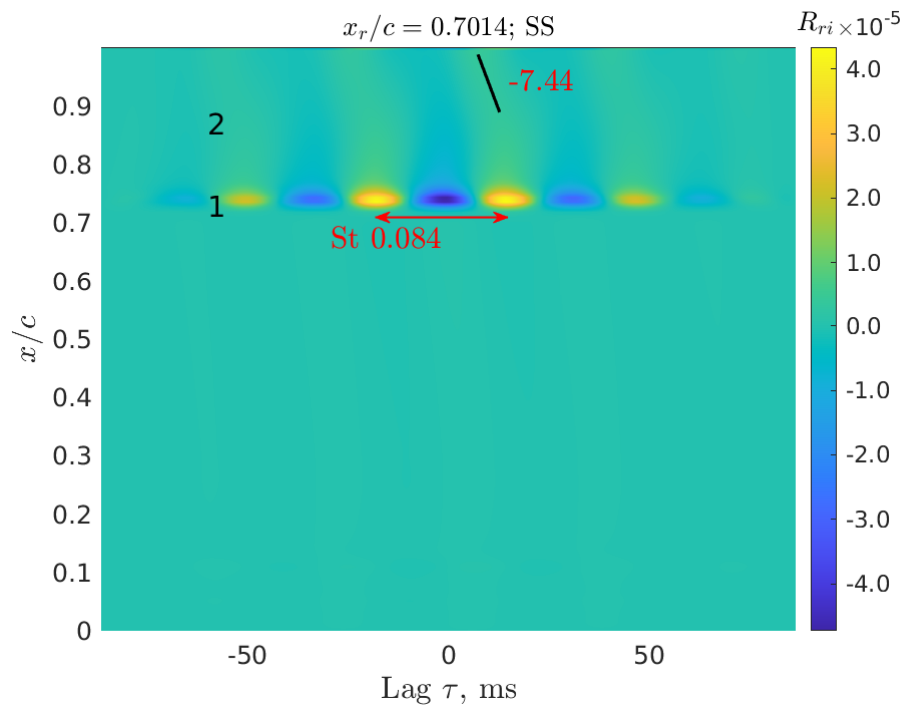}
\caption{}
\label{sfig4:corrC-1d_10ss}
\end{subfigure}
\begin{subfigure}[b]{0.49\textwidth}
\centering
\includegraphics[width=1\textwidth,keepaspectratio]{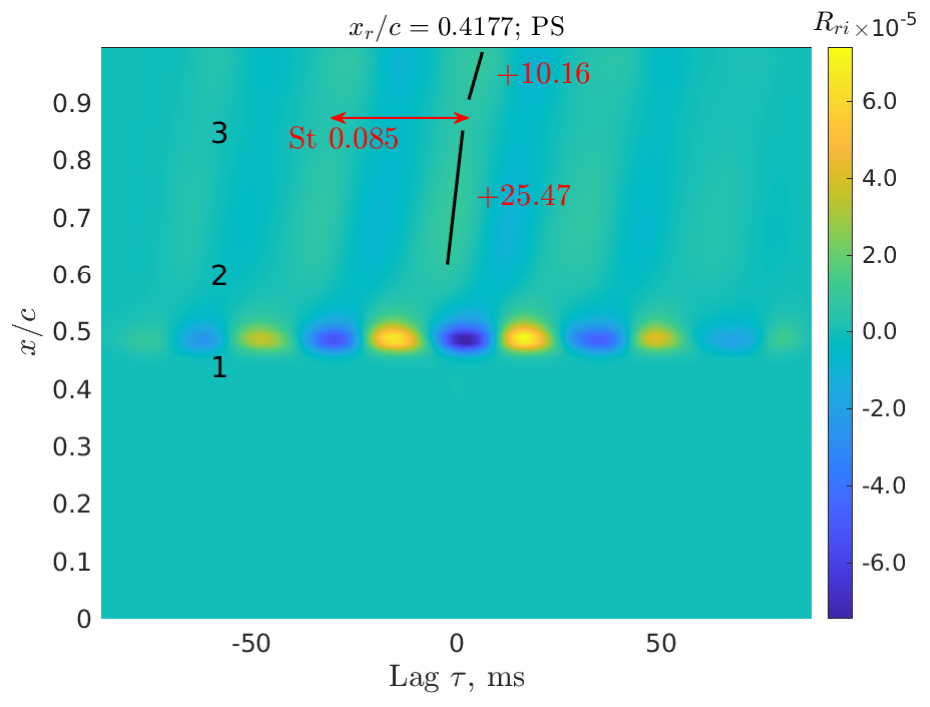}
\caption{}
\label{sfig4:corrC-1d_10ps}
\end{subfigure}
\begin{subfigure}[b]{0.49\textwidth}
\centering
\includegraphics[width=1\textwidth,keepaspectratio]{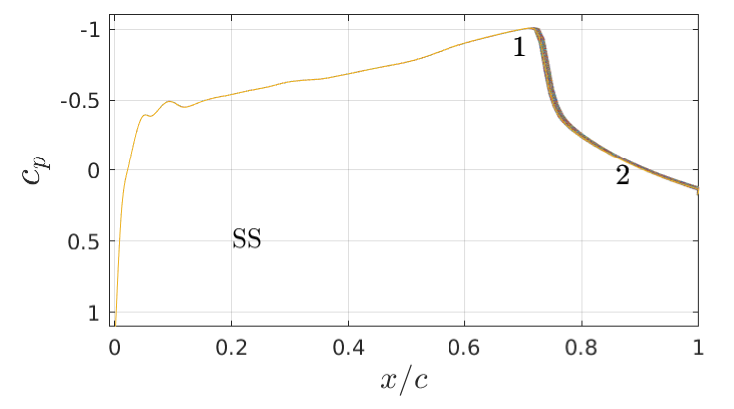}
\caption{} 
\label{sfig4:cpCFs-1d_10ss}
\end{subfigure}
\begin{subfigure}[b]{0.49\textwidth}
\centering
\includegraphics[width=1\textwidth,keepaspectratio]{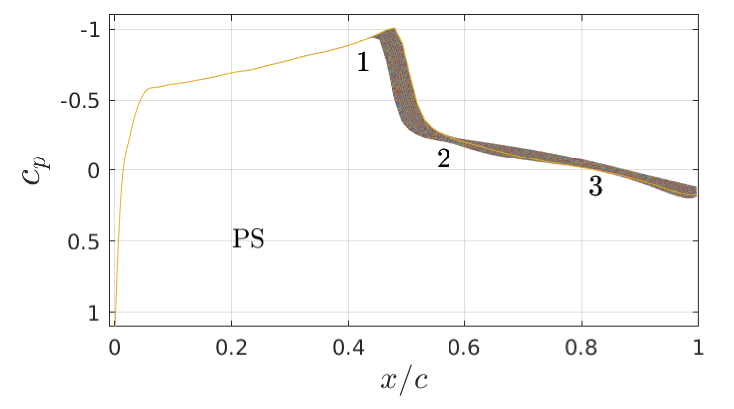}
\caption{}
\label{sfig4:cpCFs-1d_10ps}
\end{subfigure}
\caption{$-1^\circ$ AoA: Streamwise cross-correlations $R_{ri}$ (a,b) and streamwise temporal $c_p$ distributions (c,d) on the suction and pressure surfaces at $10\%$ span}
\label{fig4:corrStreamw-1d_10}
\end{figure}
 
\subsubsection{$-1^\circ$}
\label{subsec:ch4_-1d}

\begin{figure}[!ht]
\centering
\begin{subfigure}[b]{0.49\textwidth}
\centering
\includegraphics[width=1\textwidth,keepaspectratio]{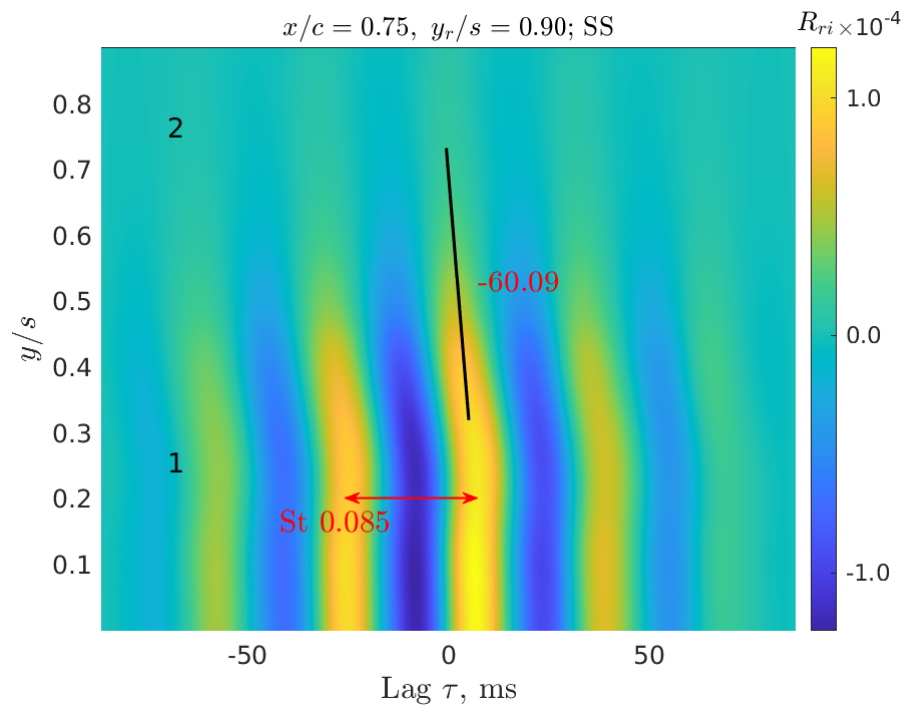}
\caption{}
\label{sfig4:corrS-1d_10ss}
\end{subfigure}
\begin{subfigure}[b]{0.49\textwidth}
\centering
\includegraphics[width=1\textwidth,keepaspectratio]{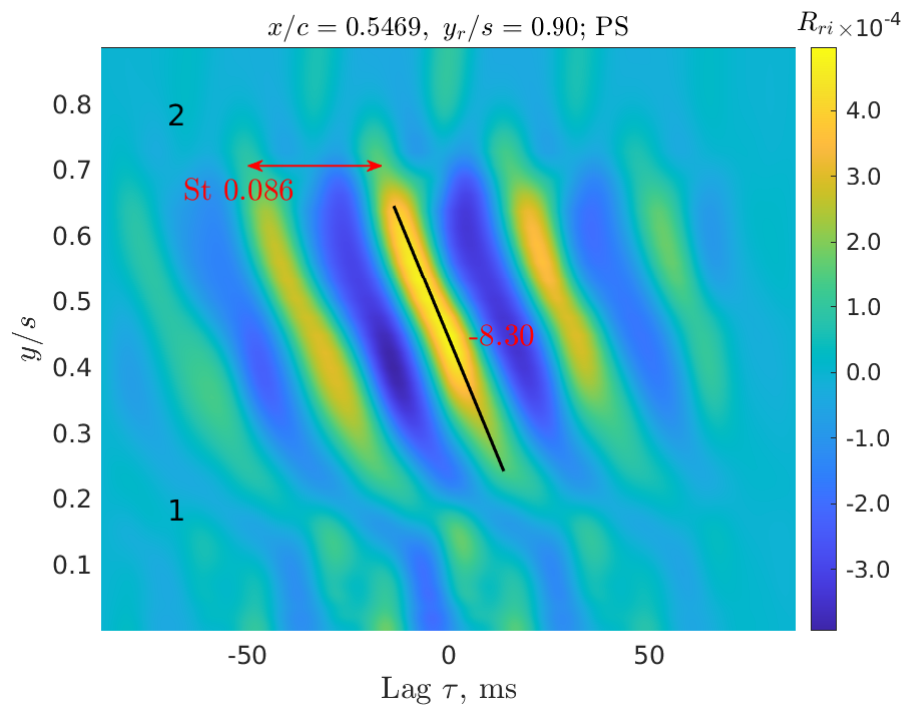}
\caption{}
\label{sfig4:corrS-1d_10ps}
\end{subfigure}
\begin{subfigure}[b]{0.49\textwidth}
\centering
\includegraphics[width=1\textwidth,keepaspectratio]{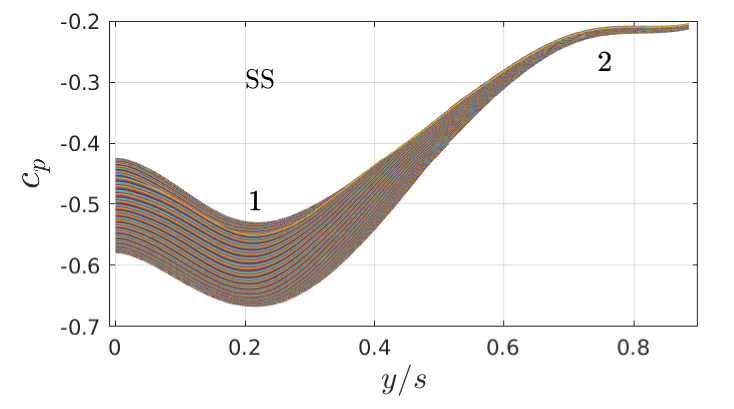}
\caption{} 
\label{sfig4:cpSFs-1d_10ss}
\end{subfigure}
\begin{subfigure}[b]{0.49\textwidth}
\centering
\includegraphics[width=1\textwidth,keepaspectratio]{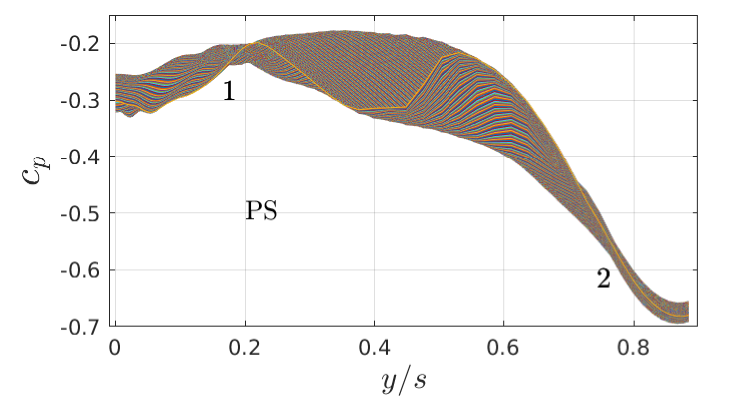}
\caption{}
\label{sfig4:cpSFs-1d_10ps}
\end{subfigure}
\begin{subfigure}{0.49\textwidth}
\centering
\includegraphics[width=\linewidth]{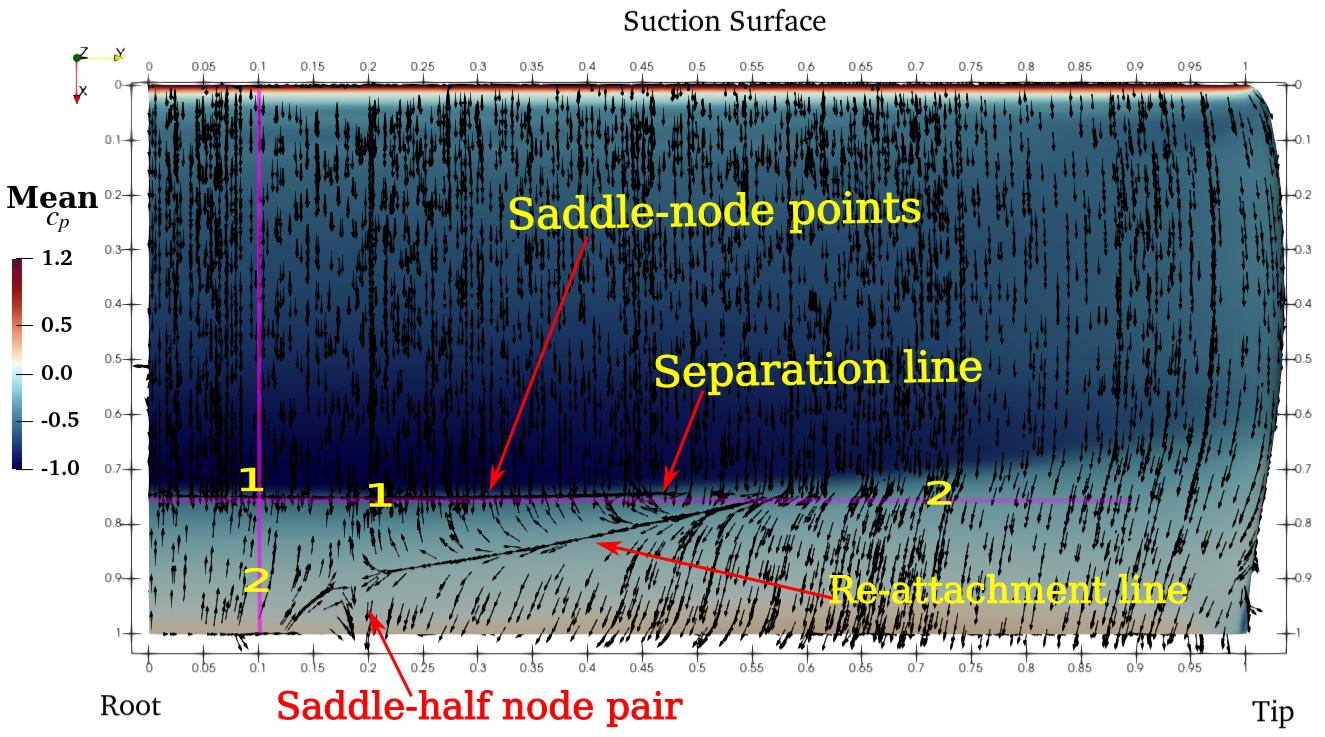}
\caption{}
\label{sfig4:msflns-1dss}
\end{subfigure}
\begin{subfigure}{0.49\textwidth}
\centering
\includegraphics[width=\linewidth]{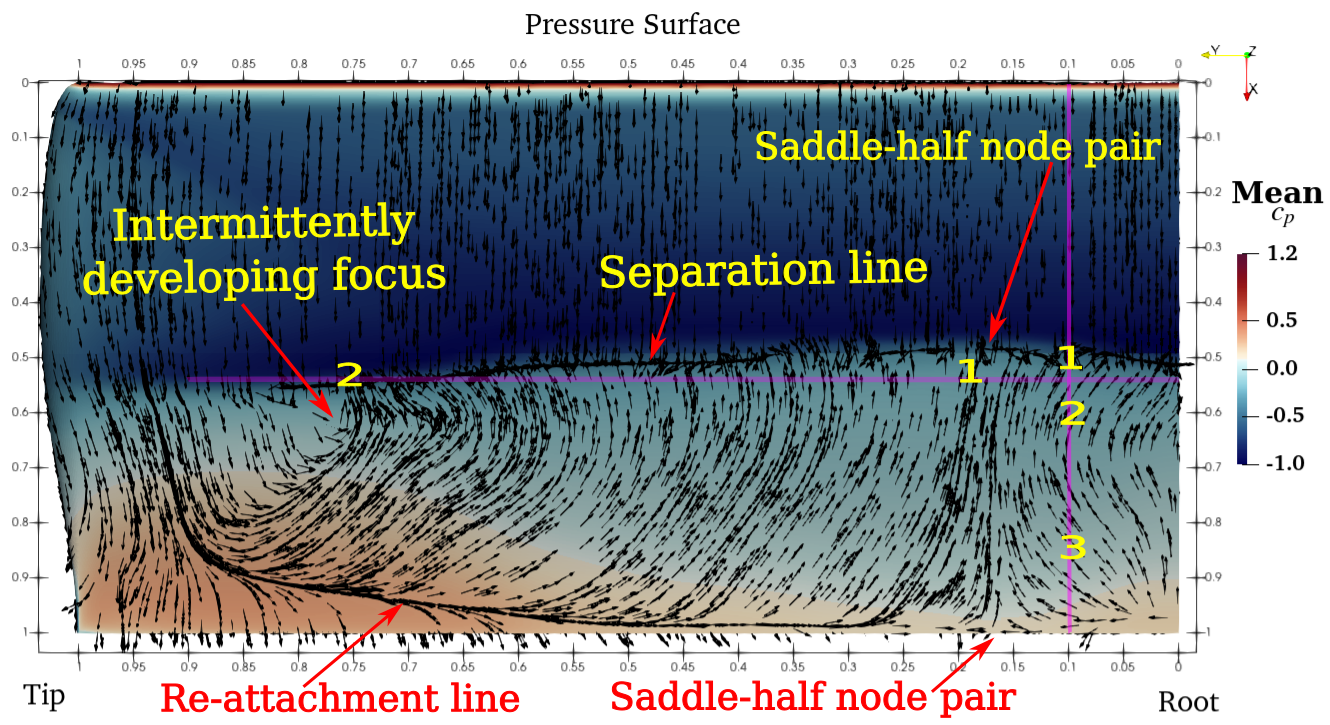}
\caption{} 
\label{sfig4:msflns-1dps}
\end{subfigure}
\caption{$-1^\circ$ AoA: Spanwise cross-correlations $R_{ri}$ (a,b), time variation of spanwise $c_p$ distributions (c,d), and mean skin-friction lines (e,f) on the suction and pressure surfaces}
\label{fig4:corrSpanw-1d_10}
\end{figure}

For this AoA, streamwise correlation and temporal $c_p$ distribution plots are presented in \autoref{fig4:corrStreamw-1d_10}; spanwise correlation and $c_p$ distribution, and mean skin-friction lines are presented in \autoref{fig4:corrSpanw-1d_10}. Spanwise correlations are of order $\mathcal{O}(1)$ higher than streamwise, implying that the spanwise wave propagation is stronger than streamwise waves.

Streamwise correlations show wave propagation only in shock excursion and flow-separated regions. On both surfaces, propagation is weaker in the flow-separated regions. However, notice speed of propagation towards the LE, in the TE region on SS, is of magnitude of order $\mathcal{O}(1)$. This propagation is hydrodynamic in nature as an incipient focus appears intermittently and disturbs the pressure field (see TE region at $10\%$ span in Supplementary Video 1A). In the shock region on both surfaces, local maxima and minima of $R_{ri}$ occur alternatively due to the to and fro movement of the shock. Thus, the time lag between two consecutive maxima or minima is the time period of shock excursion on both surfaces. The inverse of this time period is a buffet frequency. On both surfaces, this buffet frequency matches well with the frequency $St~0.084$ from PSD.

\begin{figure}[!h]
\centering
\begin{subfigure}[b]{0.49\textwidth}
\centering
\includegraphics[width=1\textwidth,keepaspectratio]{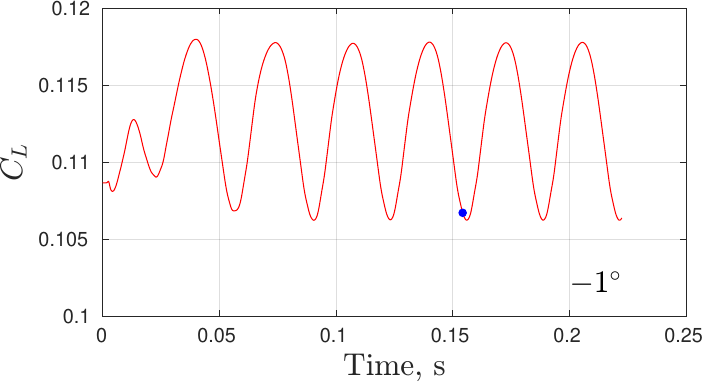}
\caption{}
\label{sfig4:ptOnCLPS0-1d}
\end{subfigure}
\begin{subfigure}[b]{0.49\textwidth}
\centering
\includegraphics[width=1\textwidth,keepaspectratio]{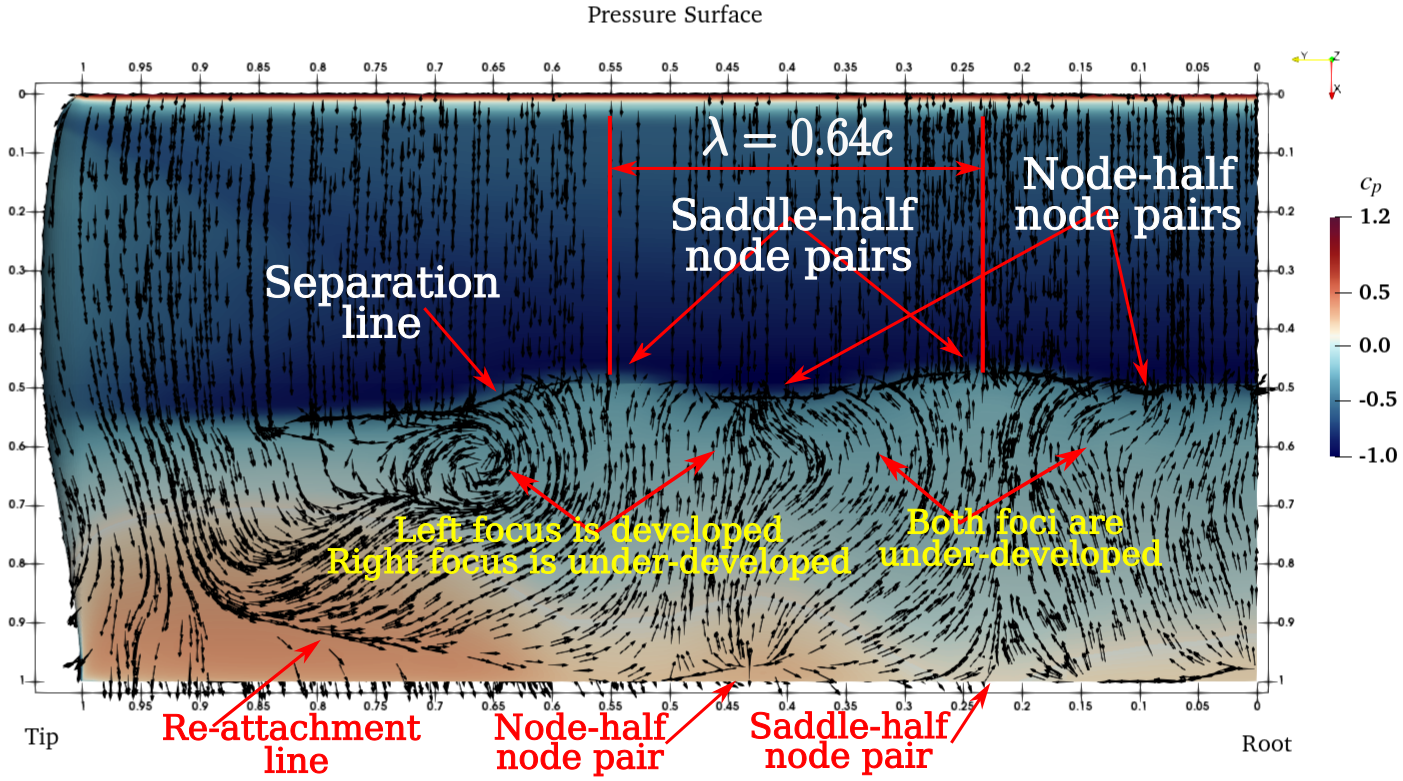}
\caption{}
\label{sfig4:waveLPS0-1d}
\end{subfigure}
\caption{$-1^\circ$ AoA: An instance on $C_L$ (a) and wavelength of propagating buffet cells at the same instance (b)}
\label{fig4:ptCLwaveLPS0-1d}
\end{figure}

In spanwise correlation on PS, inboard pressure wave between stations $1$ and $2$ propagates at velocity $V_p=-8.30$ m/s towards the wing's root. This is much lower than the inboard propagation velocity $V_p=-60.09$ m/s on SS. But, propagation frequencies are almost the same on both surfaces. So, the wavelength $\lambda_{pwp}=0.675c$ of pressure wave propagation on PS is much lower than the wavelength $5c$ on SS. The wavelength $0.675c$ on PS compares well with the measured wavelength $\lambda_{bp}=0.64c$ of buffet cells propagation on an instantaneous topology of skin-friction lines in \autoref{sfig4:waveLPS0-1d}. Propagation of pressure waves and buffet cells, both are almost in the same wavelength and propagation velocity, means very much in phase. Buffet cells propagation velocity $V_{bp}=-8.10$ m/s is calculated as $\lambda_{bp}/t_{bp}$, where $t_{bp}$ is travel time taken by a buffet cell for distance $\lambda_{bp}$. Further, notice a synchronization between inboard propagation of buffet cells and self-induced motion of contra-rotating unstable foci (see discussion of these foci in pairs in \textcolor{red}{\autoref{ssec:shockBuffMech}}) in Supplementary Video 1B. Thus, propagation of pressure perturbations and buffet cells are both carried by these 3D unstable structures. Hence, the pressure wave on PS is of a hydrodynamic nature. \autoref{sfig4:waveLPS0-1d} shows a plot of instantaneous skin-friction lines, revealing a topology of critical points along with separation and re-attachment lines. This topology is not in its fully-developed state. A fully-developed state can be observed at another instance in Supplementary Video 2B, which will resemble the general topology proposed in \autoref{ssec:shockBuffMech}. On SS, there are no buffet cells or motion of buffet cells, as seen in Supplementary Video 1A, and also, there are no fully developed foci, as seen on PS.

Mean skin-friction lines alone in \autoref{sfig4:msflns-1dss} on SS and in \autoref{sfig4:msflns-1dps} on PS do not provide a complete view of the topology of critical points. The instantaneous change of topologies on suction and pressure surfaces are noticed in Supplementary Videos 1A and 1B, respectively. Similarly, inferences from correlation, temporal $c_p$ distribution, and skin-friction lines plots are made for other AoAs.

\begin{figure}[!ht]
\centering
\begin{subfigure}[b]{0.49\textwidth}
\centering
\includegraphics[width=1\textwidth,keepaspectratio]{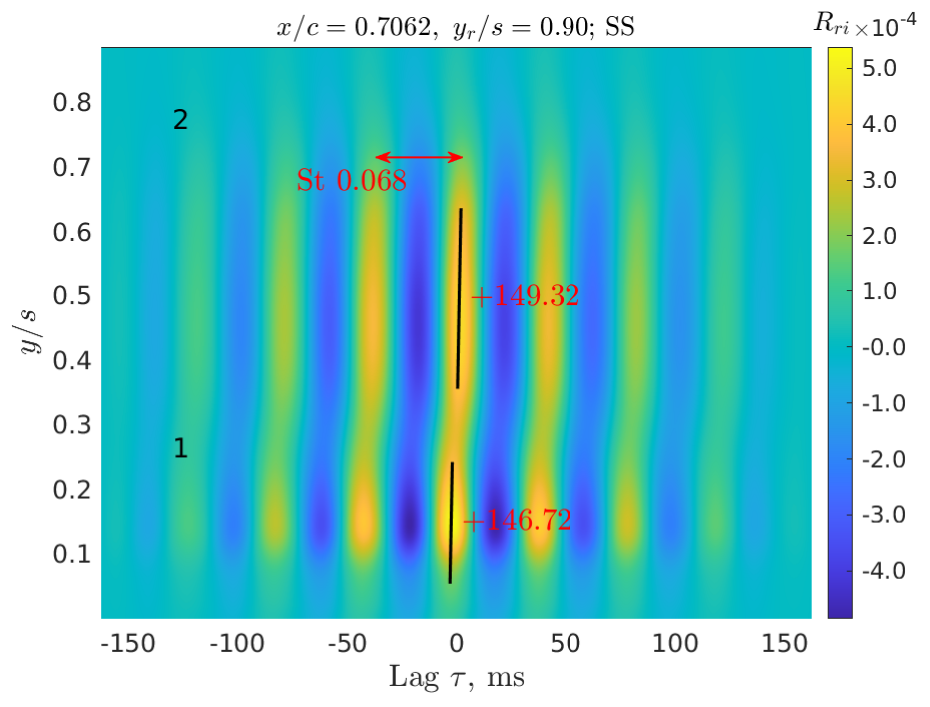}
\caption{}
\label{sfig4:corrS0d_10ss}
\end{subfigure}
\begin{subfigure}[b]{0.49\textwidth}
\centering
\includegraphics[width=1\textwidth,keepaspectratio]{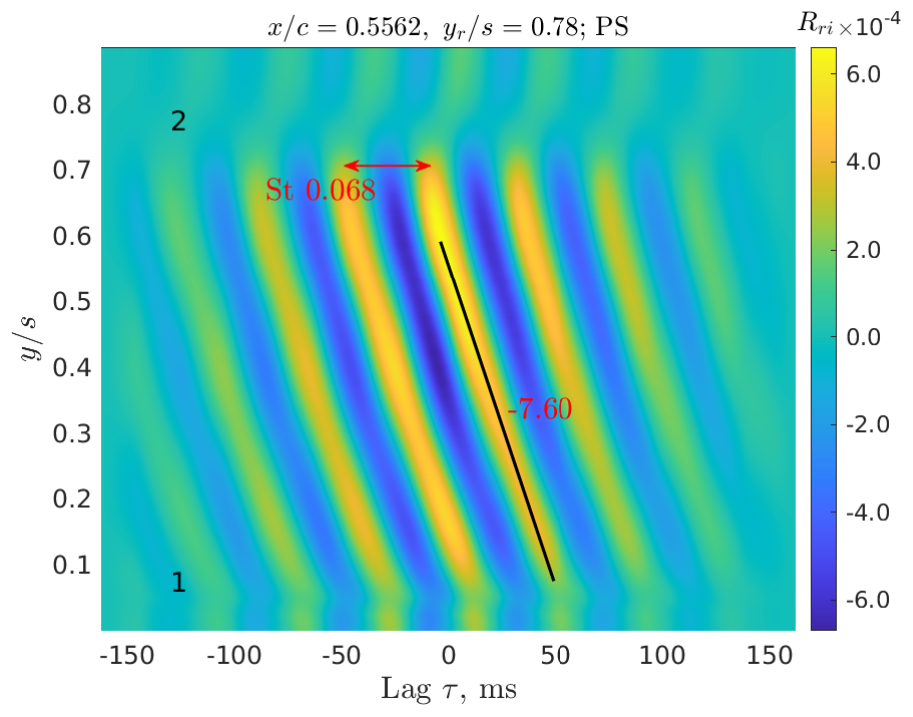}
\caption{}
\label{sfig4:corrS0d_10ps}
\end{subfigure}
\begin{subfigure}{0.49\textwidth}
\centering
\includegraphics[width=\linewidth]{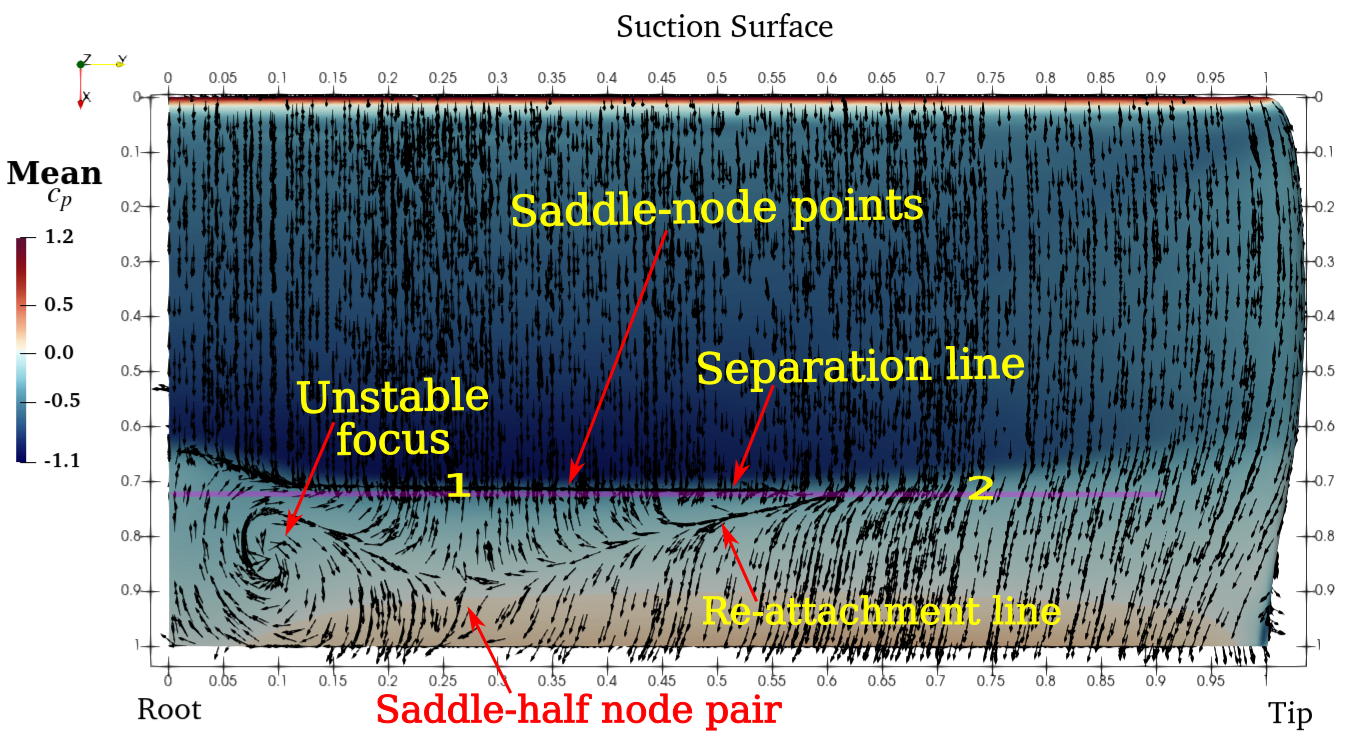}
\caption{}
\label{sfig4:msflns0dss}
\end{subfigure}
\begin{subfigure}{0.49\textwidth}
\centering
\includegraphics[width=\linewidth]{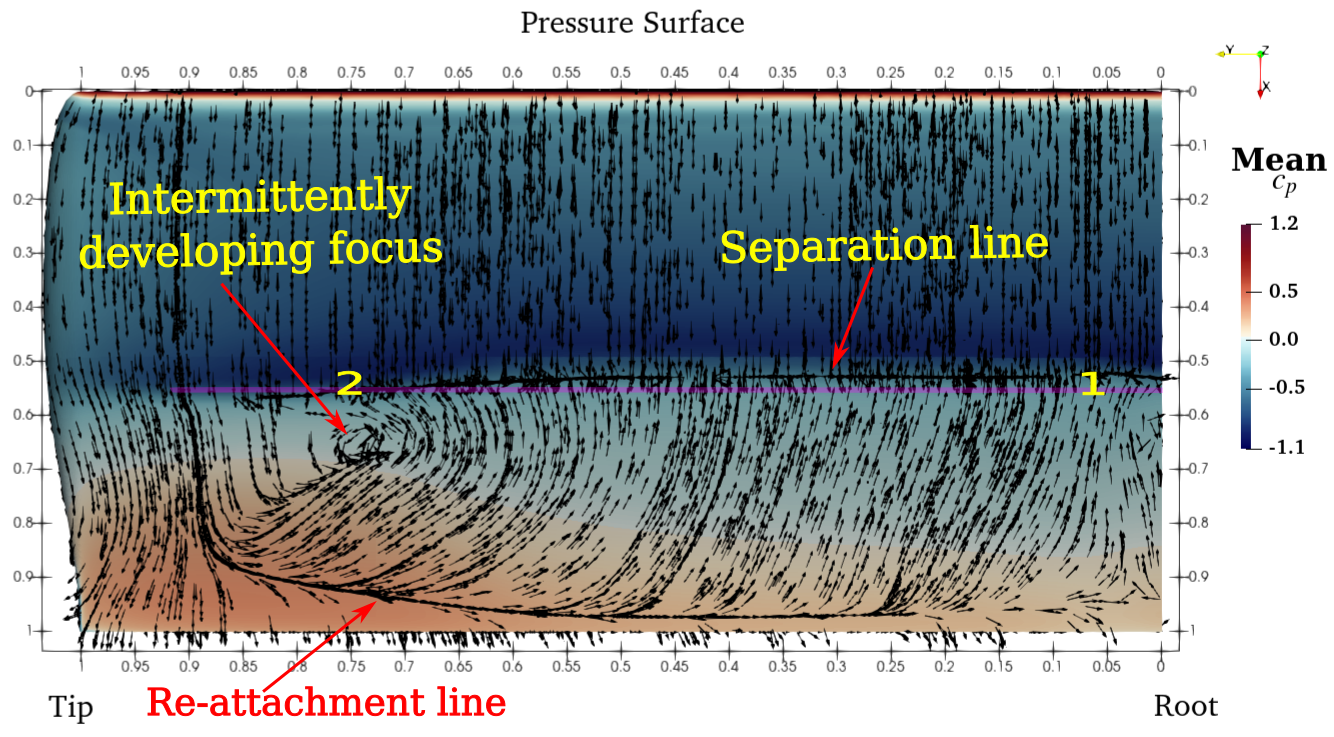}
\caption{}
\label{sfig4:msflns0dps}
\end{subfigure}
\caption{$0^\circ$ AoA: Spanwise cross-correlations $R_{ri}$ (a,b) and mean skin-friction lines (c,d) on the suction and pressure surfaces}
\label{fig4:corrSpanw0d_10}
\end{figure}

\subsubsection{$0^\circ$}
\label{subsec:ch4_0d}

\begin{figure}[!ht]
\centering
\begin{subfigure}[b]{0.49\textwidth}
\centering
\includegraphics[width=1\textwidth,keepaspectratio]{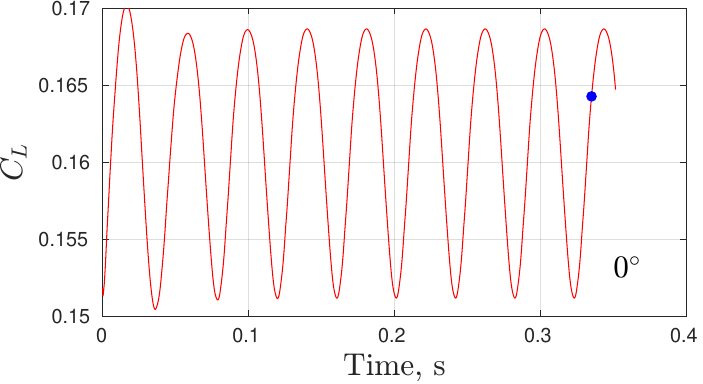}
\caption{}
\label{sfig4:ptOnCLPS0d}
\end{subfigure}
\begin{subfigure}[b]{0.49\textwidth}
\centering
\includegraphics[width=1\textwidth,keepaspectratio]{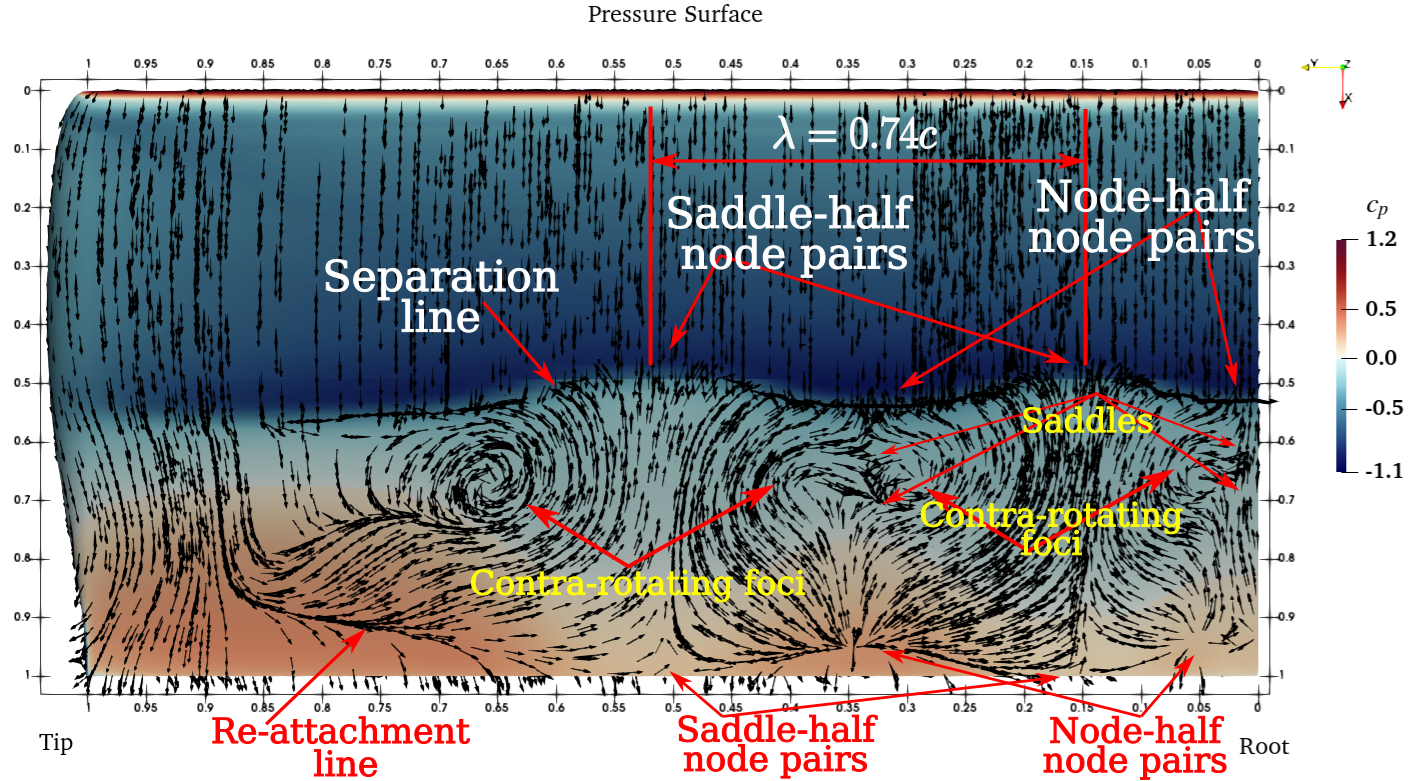}
\caption{}
\label{sfig4:waveLPS0d}
\end{subfigure}
\caption{$0^\circ$ AoA: An instance on $C_L$ (a) and wavelength of propagating buffet cells at the same instance (b)}
\label{fig4:ptCLwaveLPS0d}
\end{figure}

Spanwise correlations with mean skin-friction lines are given in \autoref{fig4:corrSpanw0d_10}. The propagation velocity of $-7.6$ m/s (hydrodynamic in nature) on PS is much lower than that of $V_p=149.32$ m/s (acoustic in nature) on SS between marks $1$ and $2$. The propagation frequency $St~0.068$ is the same on both surfaces. The wavelength $\lambda_{pwp}=0.78c$ on PS is much lower than $15.37c$ in mid-span on SS. $\lambda_{pwp}=0.78c$ on PS matches well with the wavelength $\lambda_{bp}=0.74c$ on instantaneous skin-friction lines in \autoref{fig4:ptCLwaveLPS0d}. The topology of critical points at this instance is in a developed state and resembles the general topology to much extent. This topology contains a sum of $8$ numbers of $4$ foci and $4$ nodes, which is equal to $8$ numbers of saddle points in the separated region. Thus, the topology satisfies the relation in \autoref{eq:saddle_nodesfoci_rel}.

On SS, Supplementary Video 2A shows no pair of contra-rotating unstable foci; hence, no buffet cell is visible. But animation shows only a single clockwise rotating unstable focus in its fully-grown state in near-root region. This focus is also visible in mean skin-friction lines in \autoref{sfig4:msflns0dss}. The reverse flow caused by this focus creates a curve on separation line near the root. This focus oscillates in both streamwise and spanwise directions, and disturbs pressure distribution in both directions.

\begin{figure}[!ht]
\centering
\begin{subfigure}[b]{0.49\textwidth}
\centering
\includegraphics[width=1\textwidth,keepaspectratio]{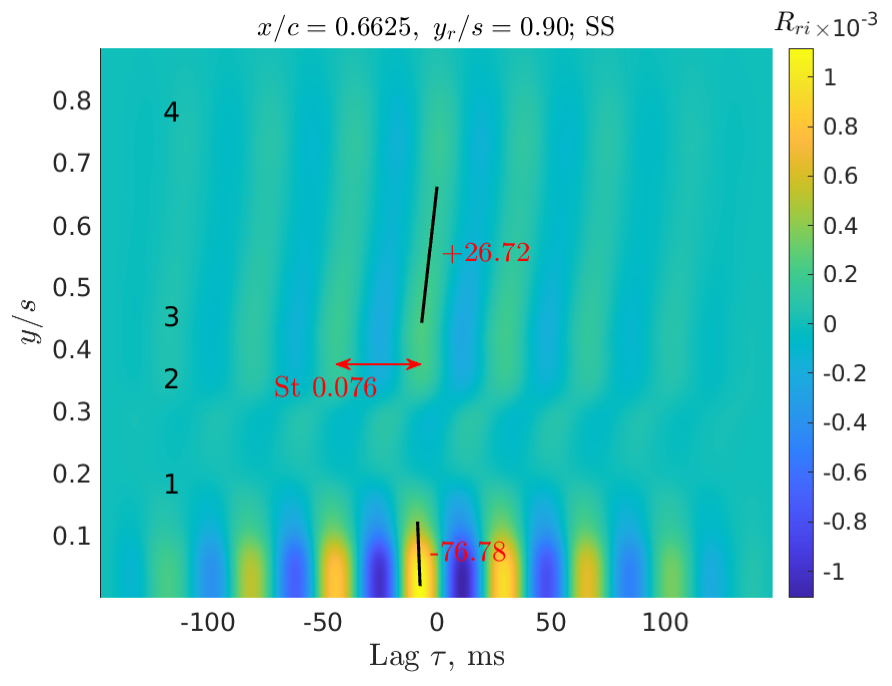}
\caption{}
\label{sfig4:corrS1d_10ss}
\end{subfigure}
\begin{subfigure}[b]{0.49\textwidth}
\centering
\includegraphics[width=1\textwidth,keepaspectratio]{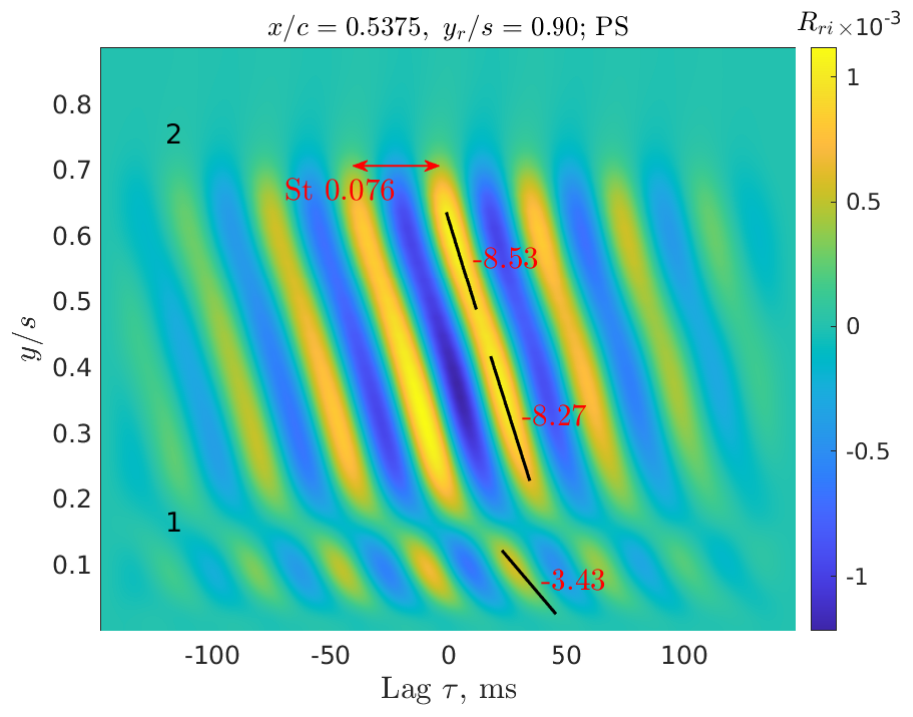}
\caption{}
\label{sfig4:corrS1d_10ps}
\end{subfigure}
\begin{subfigure}{0.49\textwidth}
\centering
\includegraphics[width=\linewidth]{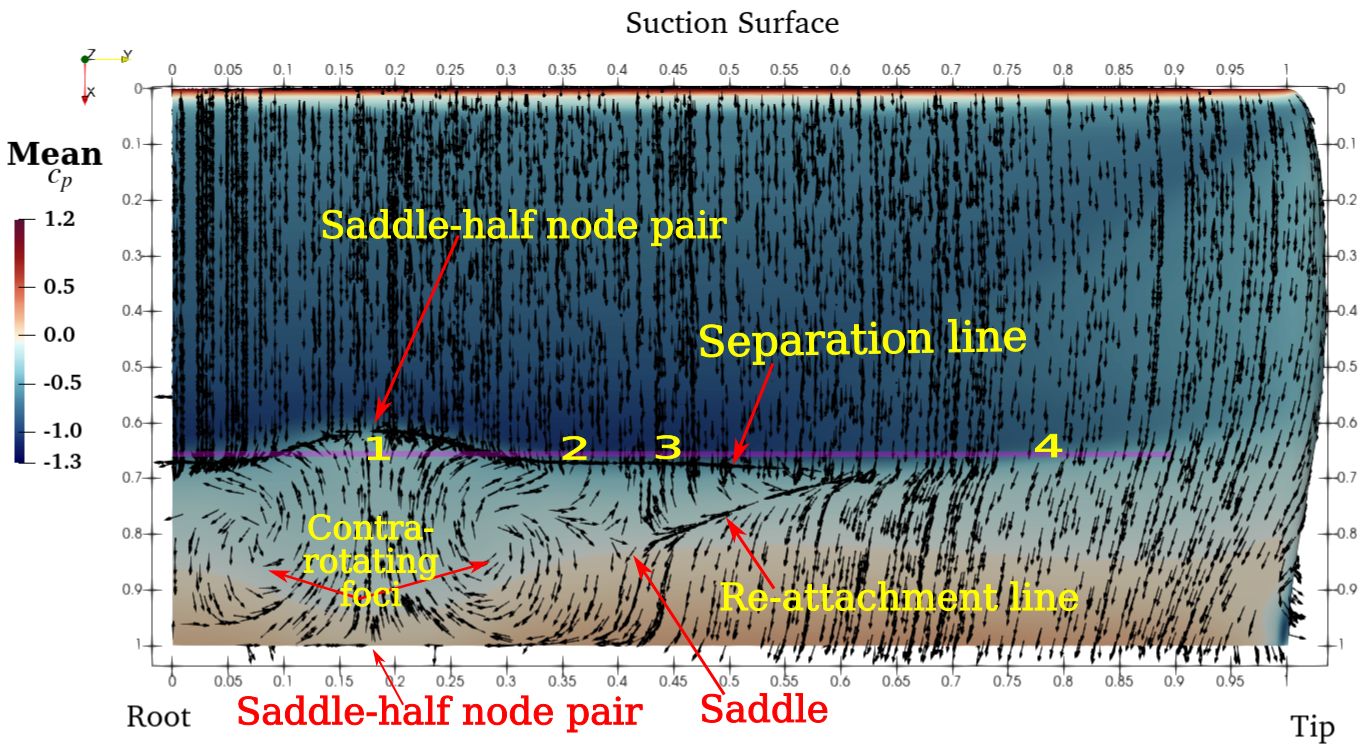}
\caption{}
\label{sfig4:msflns1dss}
\end{subfigure}
\begin{subfigure}{0.49\textwidth}
\centering
\includegraphics[width=\linewidth]{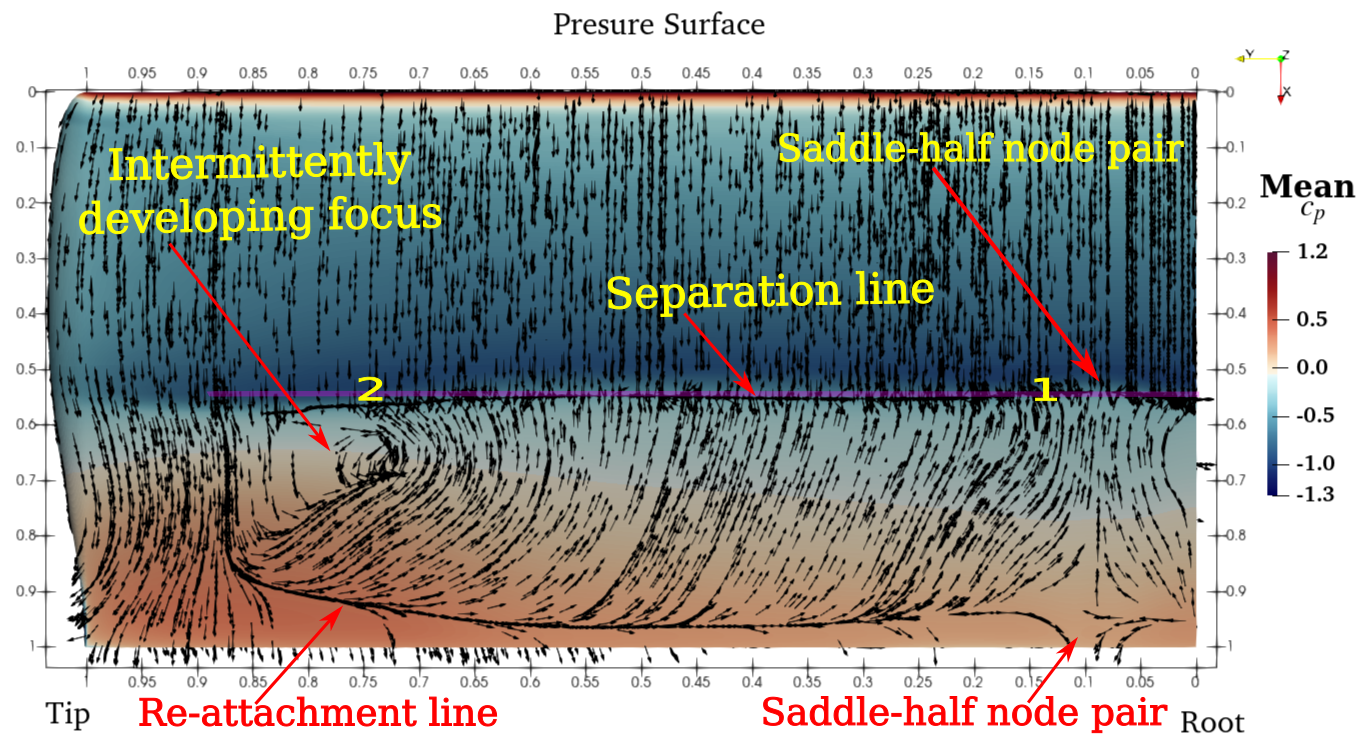}
\caption{} 
\label{sfig4:msflns1dps}
\end{subfigure}
\caption{$1^\circ$ AoA: Spanwise cross-correlations $R_{ri}$ (a,b) and mean skin-friction lines (c,d) on the suction and pressure surfaces}
\label{fig4:corrSpanw1d_10}
\end{figure}

\subsubsection{$1^\circ$}
\label{subsec:ch4_1d}

\begin{figure}[!ht]
\centering
\begin{subfigure}[b]{0.49\textwidth}
\centering
\includegraphics[width=1\textwidth,keepaspectratio]{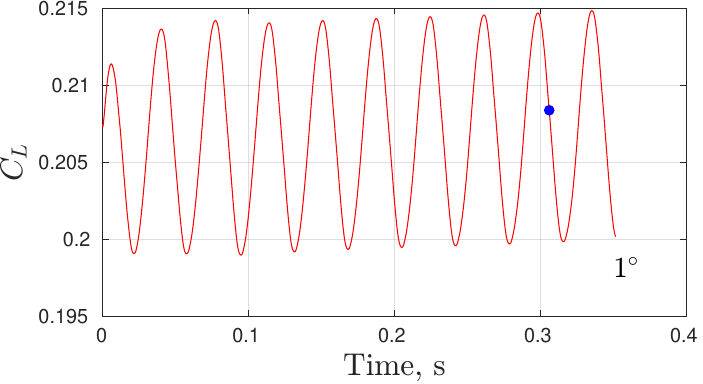}
\caption{}
\label{sfig4:ptOnCLPS1d}
\end{subfigure}
\begin{subfigure}[b]{0.49\textwidth}
\centering
\includegraphics[width=1\textwidth,keepaspectratio]{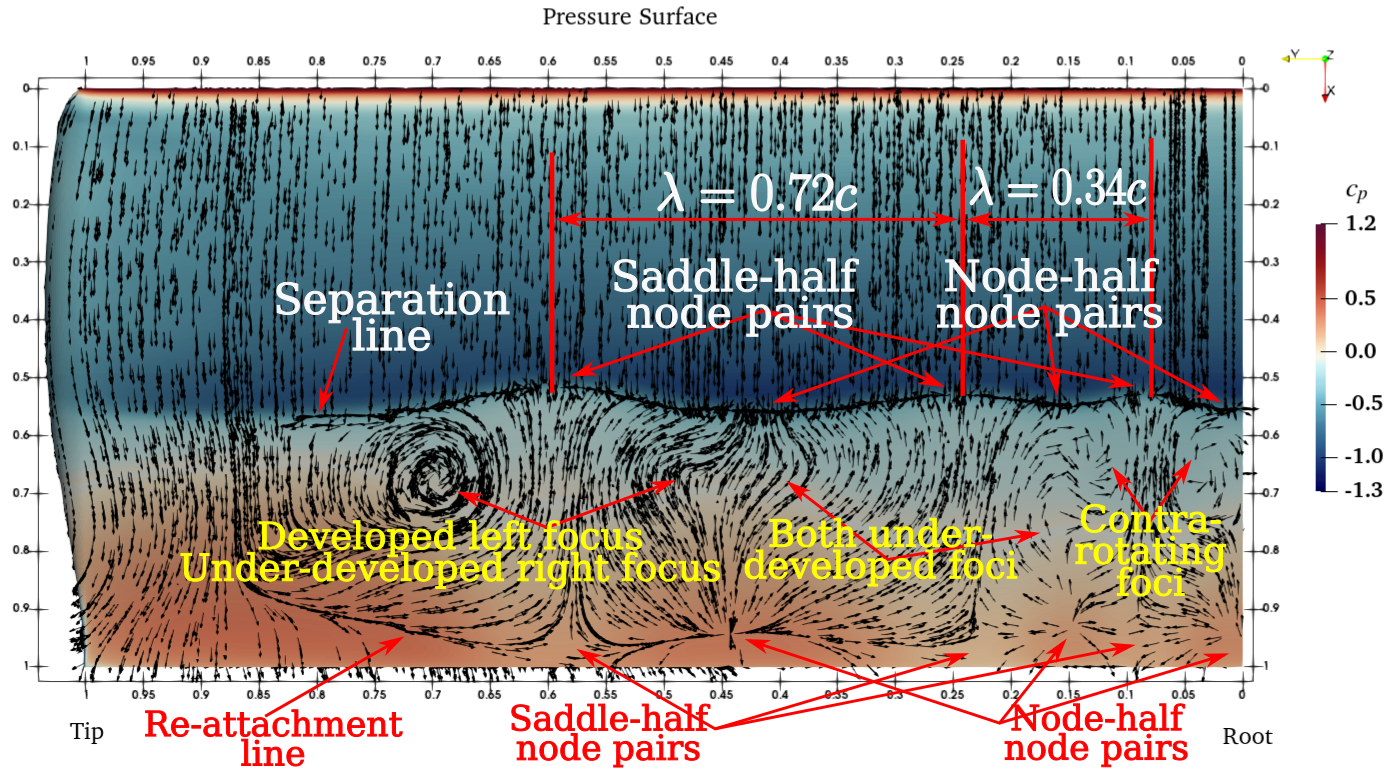}
\caption{}
\label{sfig4:waveLPS1d}
\end{subfigure}
\caption{$1^\circ$ AoA: An instance on $C_L$ (a) and wavelength of propagating buffet cells at the same instance (b)}
\label{fig4:ptCLwaveLPS1d}
\end{figure}

Spanwise correlations and mean skin-friction lines are arranged together in \autoref{fig4:corrSpanw1d_10}. Pressure correlation is comparatively strong near the wing's root on SS. The first three points on SS are of specific locations on skin-friction lines. $1$ is at $\approx0.17s$, which is a spanwise location of saddle-half node on separation line in \autoref{sfig4:msflns1dss}. $2$ and $3$ are the left and right points about intermittently developing saddle. On PS, inboard propagation average velocity is $-8.4$ m/s between $1$ and $2$. There is also inboard propagation velocity $-3.43$ m/s near the root, inboard of $1$. Typically, two different sizes of buffet cells can be observed in Supplementary Video 3B. Larger buffet cells are in mid-span associated with $-8.4$ m/s and smaller in near-root region with $-3.43$ m/s. Both are propagating with frequency $St~0.076$. Wavelength $\lambda_{pwp}=0.77c$ is computed from velocity $-8.4$ m/s and $\lambda_{pwp}=0.32c$ from $-3.43$ m/s. $\lambda_{pwp}=0.77c$ and $\lambda_{pwp}=0.32c$ compare well with the corresponding $\lambda_{bp}=0.72c$ and $\lambda_{bp}=0.34c$ on skin-friction lines in \autoref{fig4:ptCLwaveLPS1d}. This topology is an underdeveloped state and does not completely resemble the general topology. Here, large and small sizes of buffet cells are present simultaneously on PS.

There are only spanwise oscillations of a buffet cell in the root region on SS, as seen clearly in Supplementary Video 3A. The frequency of buffet cell oscillation is the shock oscillation frequency of $St~0.076$ at $10\%$ span as they are exactly in phase. Notice a systematic development of a buffet cell on SS in near-root region from $-1^\circ$ to $1^\circ$ AoA in the corresponding supplementary videos. There is an incipient focus at $-1^\circ$, which turns out to be a fully-grown unstable focus at $0^\circ$, then it drives the reverse flow against the separation line with another counter-rotating focus at $1^\circ$. The curvature formed on separation line by such a pair is called a buffet cell. The same is discussed in detail in \autoref{ssec:shockBuffMech}.

\begin{figure}[!ht]
\centering
\begin{subfigure}[b]{0.49\textwidth}
\centering
\includegraphics[width=1\textwidth,keepaspectratio]{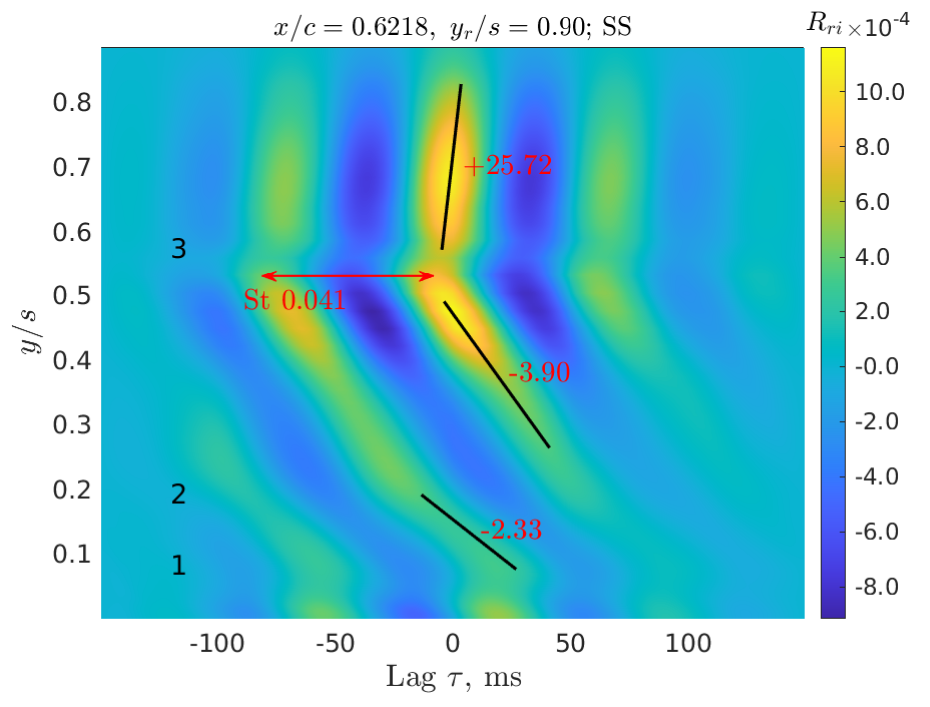}
\caption{}
\label{sfig4:corrS3d_10ss}
\end{subfigure}
\begin{subfigure}[b]{0.49\textwidth}
\centering
\includegraphics[width=1\textwidth,keepaspectratio]{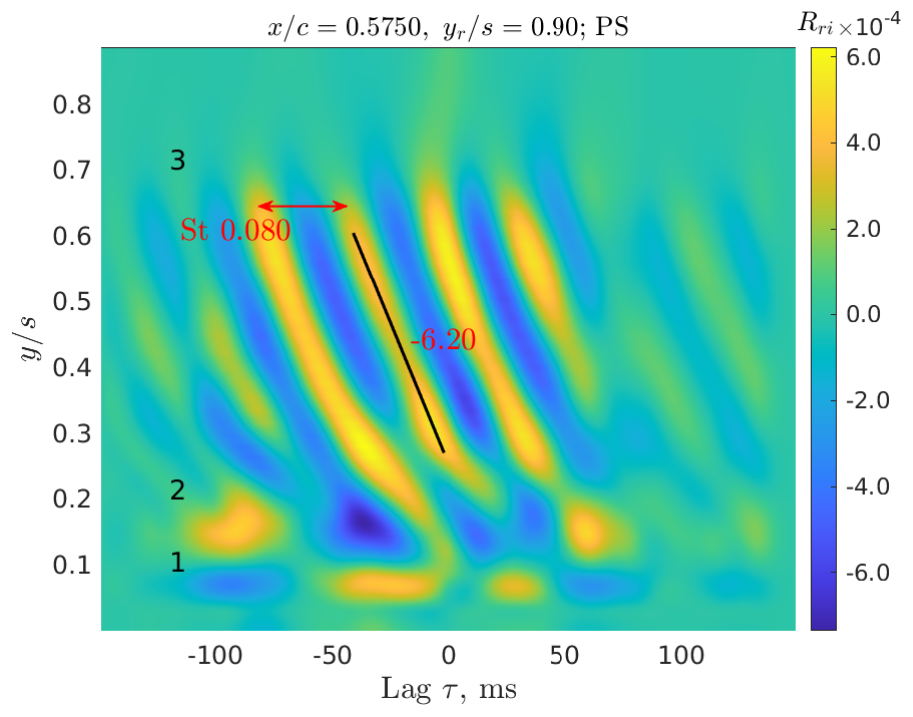}
\caption{}
\label{sfig4:corrS3d_10ps}
\end{subfigure}
\begin{subfigure}{0.49\textwidth}
\centering
\includegraphics[width=\linewidth]{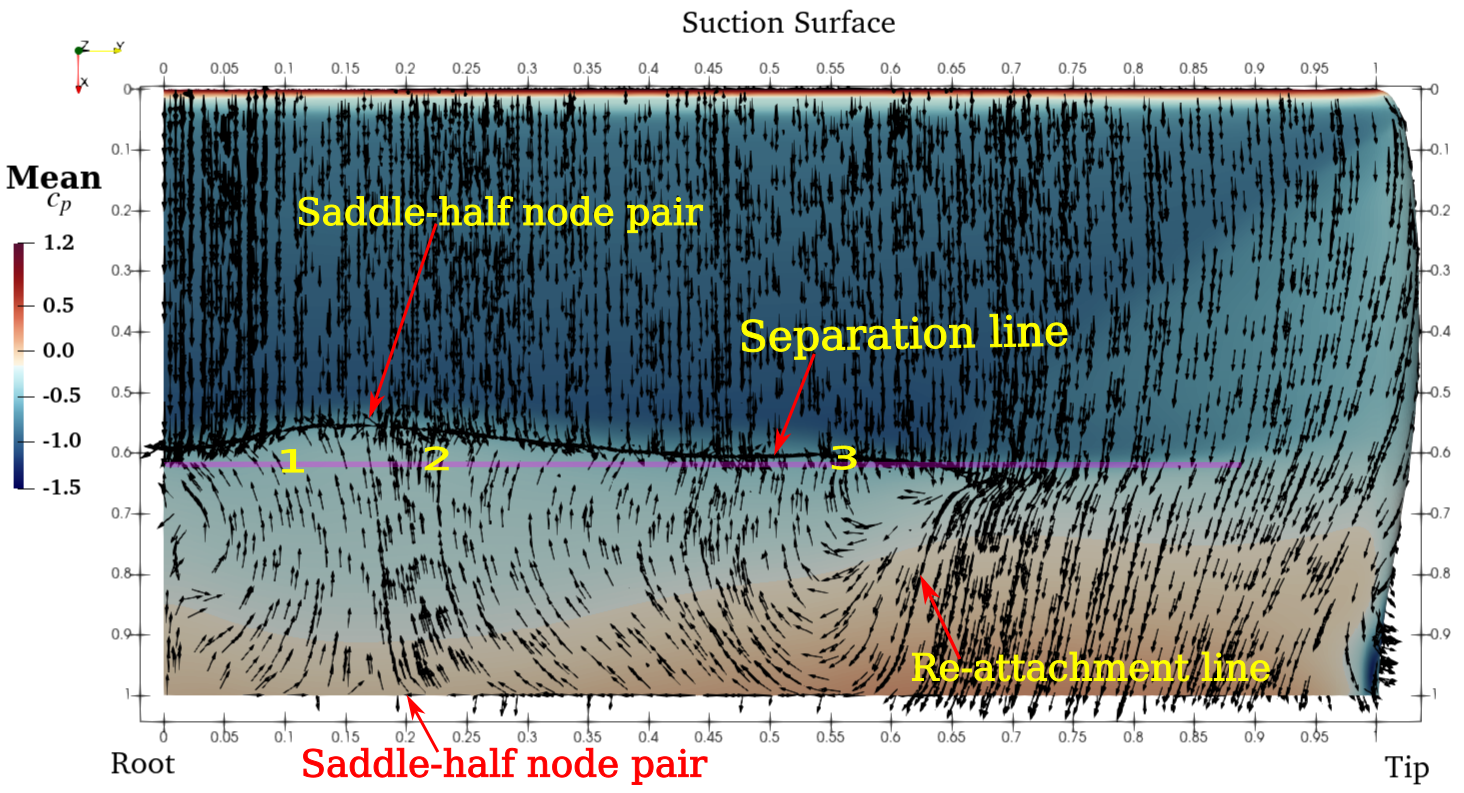}
\caption{}
\label{sfig4:msflns3dss}
\end{subfigure}
\begin{subfigure}{0.49\textwidth}
\centering
\includegraphics[width=\linewidth]{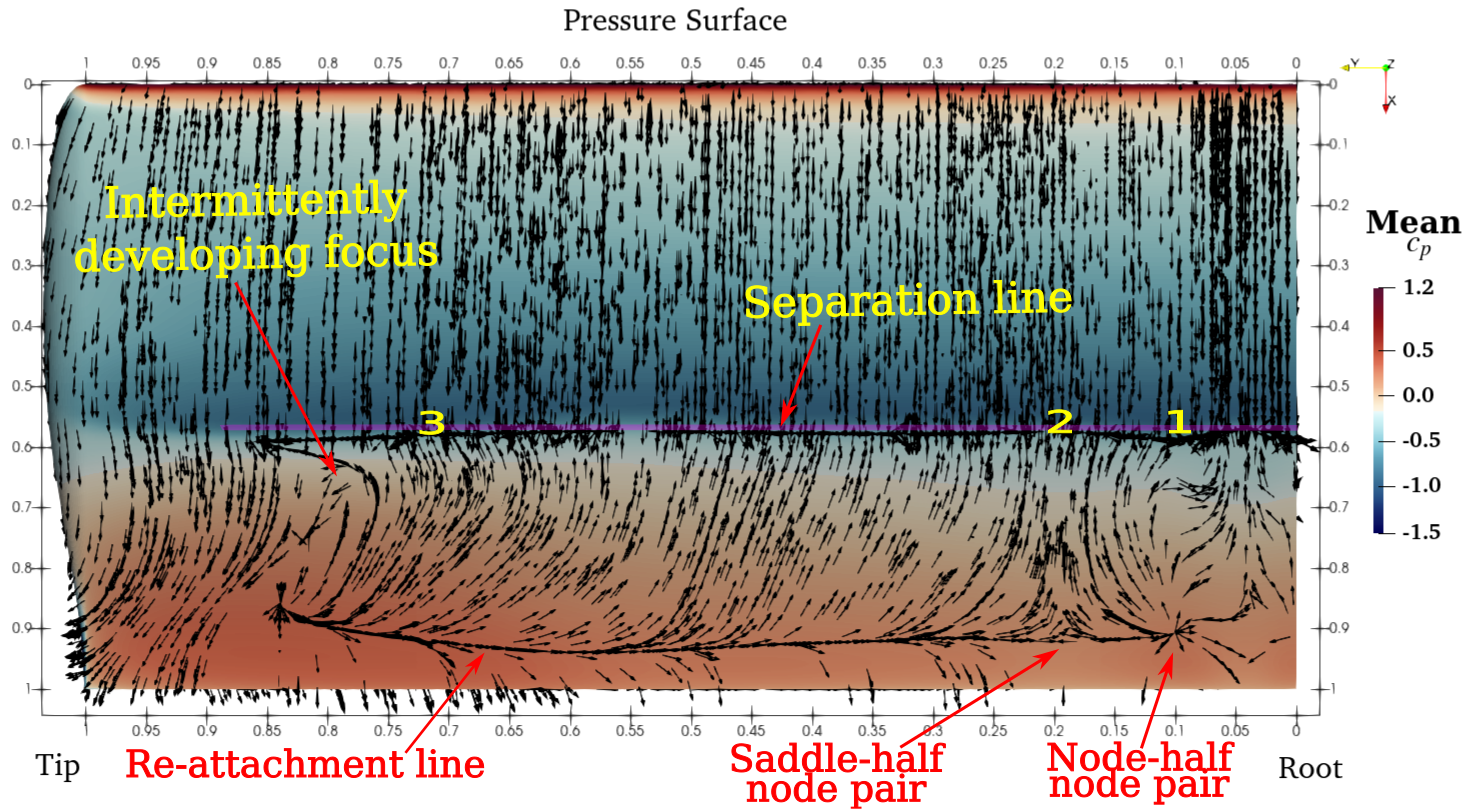}
\caption{} \label{sfig4:msflns3dps}
\end{subfigure}
\caption{$3^\circ$ AoA: Spanwise cross-correlations $R_{ri}$ (a,b) and mean skin-friction lines (c,d) on the suction and pressure surfaces}
\label{fig4:corrSpanw3d_10}
\end{figure}

\subsubsection{$3^\circ$}
\label{subsec:ch4_3d}

\begin{figure}[!ht]
\centering
\begin{subfigure}[b]{0.49\textwidth}
\centering
\includegraphics[width=1\textwidth,keepaspectratio]{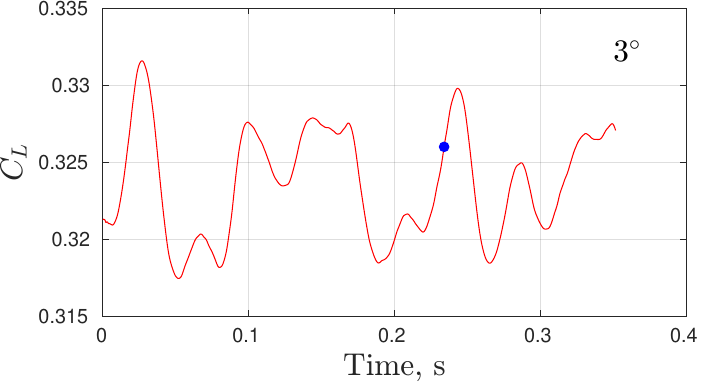}
\caption{}
\label{sfig4:ptOnCLSS3d}
\end{subfigure}
\begin{subfigure}[b]{0.49\textwidth}
\centering
\includegraphics[width=1\textwidth,keepaspectratio]{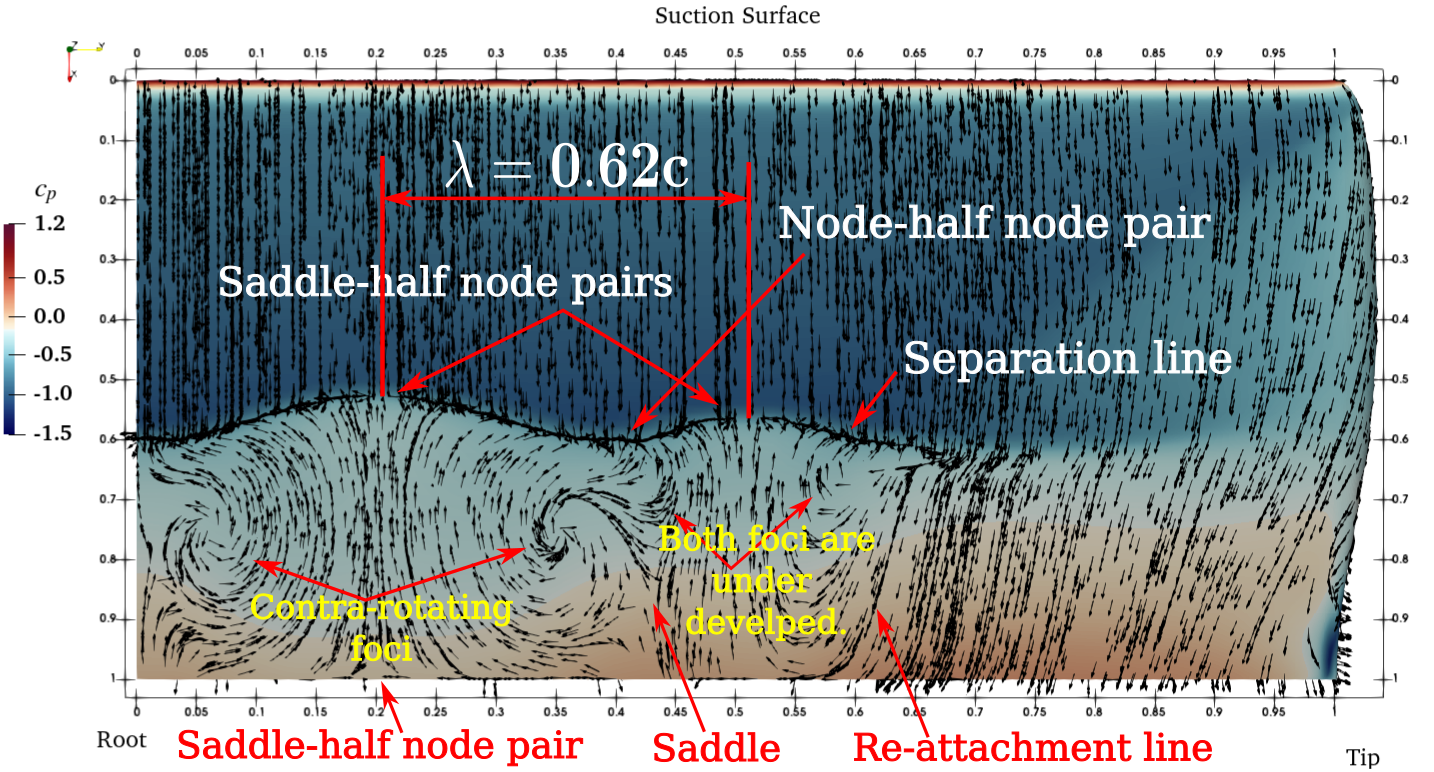}
\caption{}
\label{sfig4:waveLSS3d}
\end{subfigure}
\caption{$3^\circ$ AoA: An instance on $C_L$ (a) and wavelength of propagating buffet cells at the same instance (b)}
\label{fig4:ptCLwaveLSS3d}
\end{figure}

\begin{figure}[!ht]
\centering
\begin{subfigure}[b]{0.49\textwidth}
\centering
\includegraphics[width=1\textwidth,keepaspectratio]{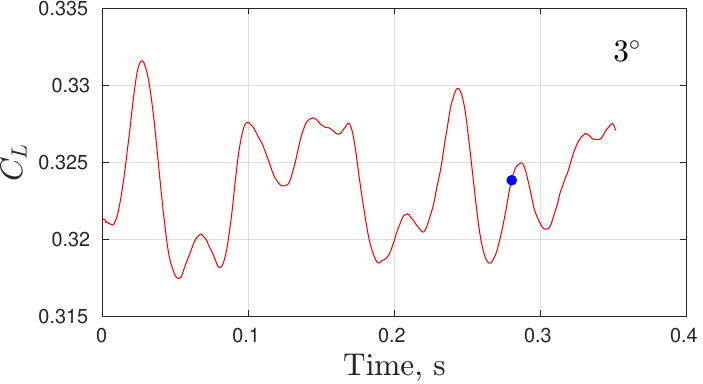}
\caption{}
\label{sfig4:ptOnCL3dSmlLrgP}
\end{subfigure}
\begin{subfigure}[b]{0.49\textwidth}
\centering
\includegraphics[width=1\textwidth,keepaspectratio]{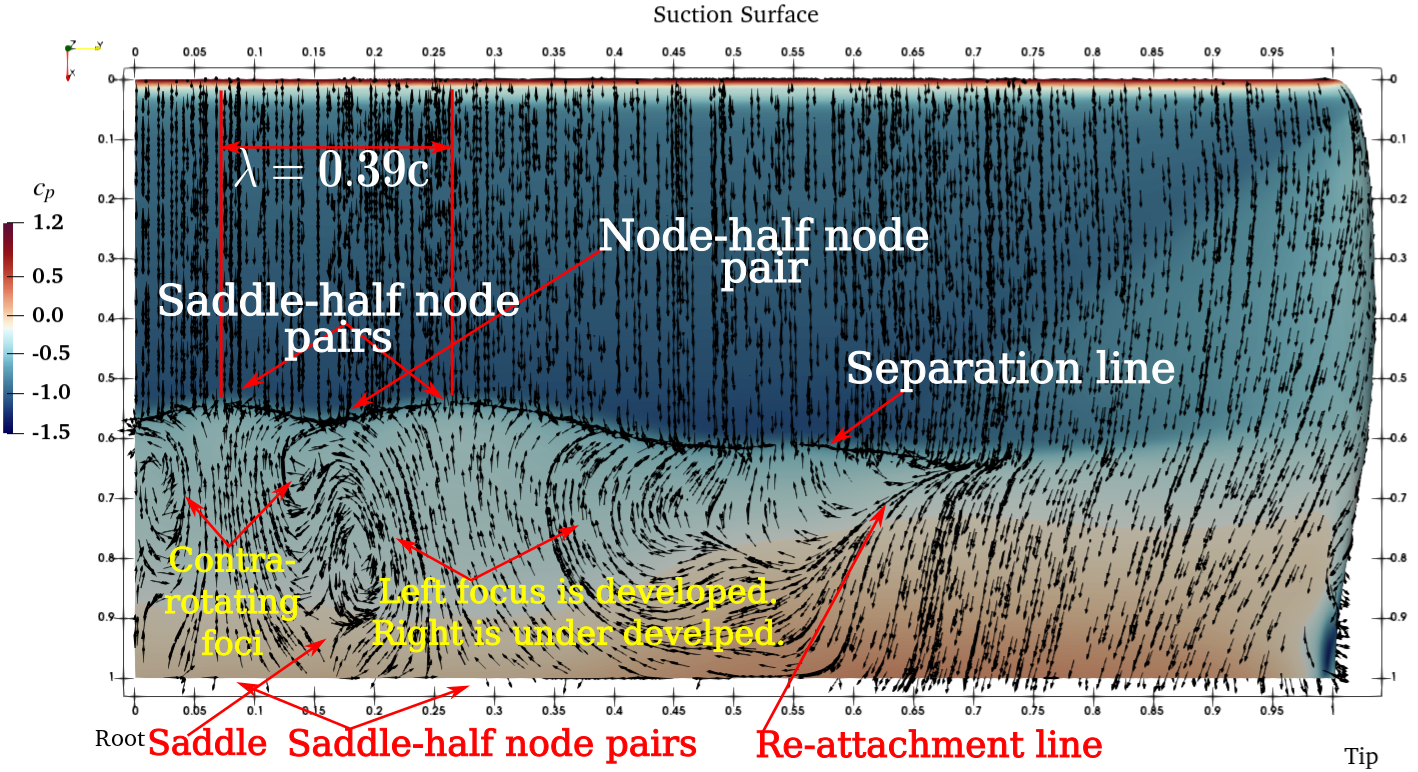}
\caption{}
\label{sfig4:smlLrgPairsSS3d}
\end{subfigure}
\caption{$3^\circ$ AoA: An instance on $C_L$ (a) and wavelength of propagating buffet cells at the same instance (b)}
\label{fig4:smlLrgPairs}
\end{figure}

Spanwise correlations and mean skin-friction lines are plotted in \autoref{fig4:corrSpanw3d_10}. On SS, inboard propagation velocity $-3.90$ m/s is from span location $3$ to $2$, and $-2.33$ m/s from $2$ to $1$ with frequency $St~0.041$. Here, propagation slows, and foci become smaller in size while reaching the root from the mid-span region, as seen in Supplementary Video 4A. That is why bigger and smaller foci are at two different instances, unlike these foci at the same instance on PS for $1^\circ$ AoA. Notice that mark $3$ is the location from where inboard hydrodynamic ($-3.9$ m/s) and outboard acoustic ($25.72$ m/s) waves originate. Wavelengths $\lambda_{pwp}=0.66c$ in mid-span and $\lambda_{pwp}=0.38c$ in near-root regions compare well with $\lambda_{bp}=0.62c$ and $\lambda_{bp}=0.39c$, respectively. The wavelength $\lambda_{bp}=0.62c$ is shown at an instance in \autoref{fig4:ptCLwaveLSS3d} and $\lambda_{bp}=0.39c$ in \autoref{fig4:smlLrgPairs}. Topologies of skin-friction lines in these instances are not in fully developed states. 

\begin{figure}[!ht]
\centering
\begin{subfigure}[b]{0.49\textwidth}
\centering
\includegraphics[width=1\textwidth,keepaspectratio]{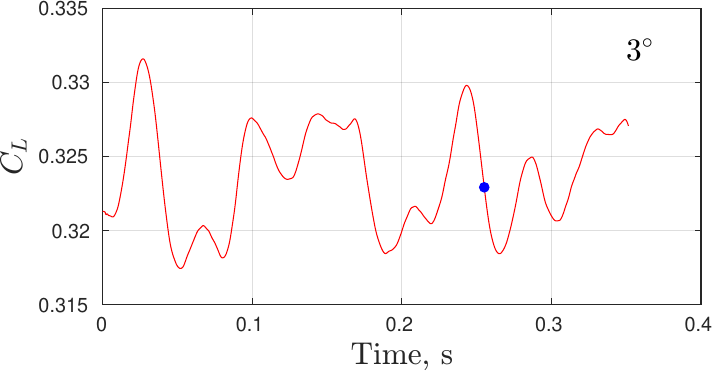}
\caption{}
\label{sfig4:ptOnCLPS3d}
\end{subfigure}
\begin{subfigure}[b]{0.49\textwidth}
\centering
\includegraphics[width=1\textwidth,keepaspectratio]{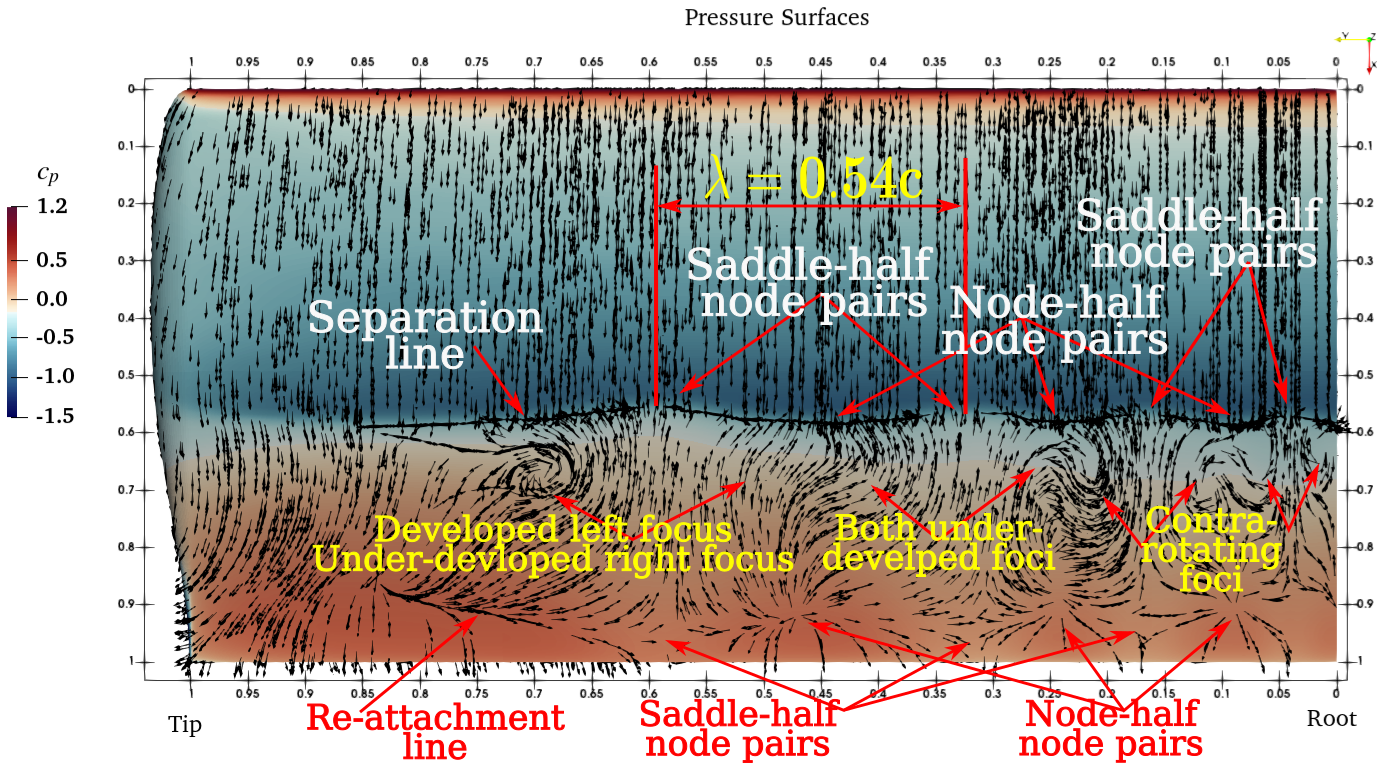}
\caption{}
\label{sfig4:waveLPS3d}
\end{subfigure}
\caption{$3^\circ$ AoA: An instance on $C_L$ (a) and wavelength of propagating buffet cells at the same instance (b)}
\label{fig4:ptCLwaveLPS3d}
\end{figure}

On PS, inboard propagation occurs at velocity $-6.20$ m/s from location $3$ to $2$ with frequency $St~0.080$. $\lambda_{pwp}=0.542c$ matches very well with $\lambda_{bp}=0.54c$ on skin-friction lines at an instance in \autoref{fig4:ptCLwaveLPS3d}. Only for this AoA, dominant frequencies are different on both surfaces, double on PS than that on SS. Both frequencies are endpoints of the band $St$ $0.04-0.08$ from PSD in \autoref{subsec:fa} for this AoA. And, inboard propagations occur on both surfaces especially for this AoA.

\begin{figure}[!ht]
\centering
\begin{subfigure}[b]{0.49\textwidth}
\centering
\includegraphics[width=1\textwidth,keepaspectratio]{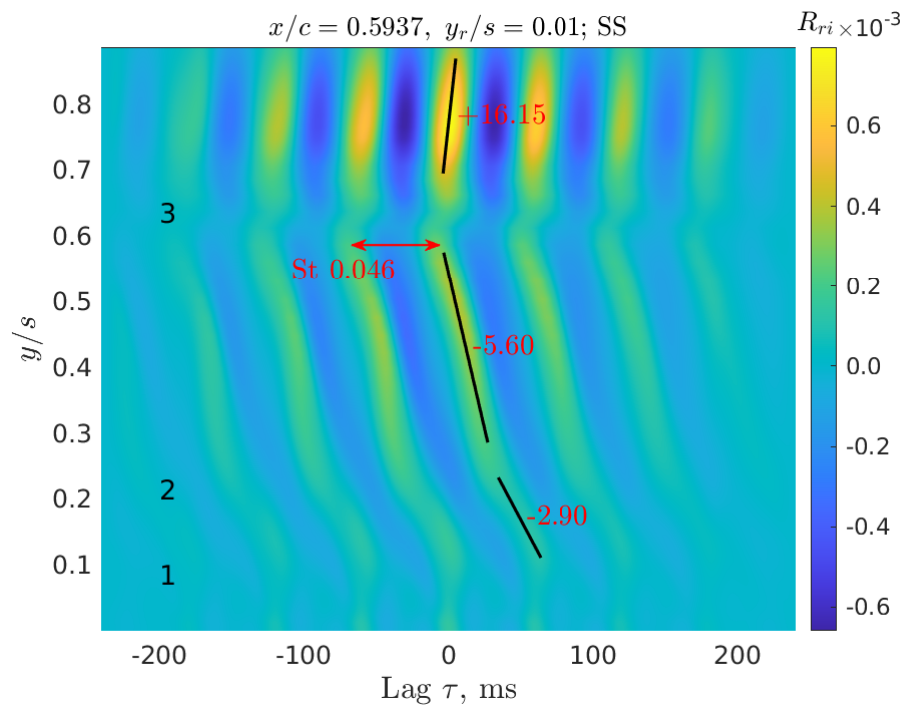}
\caption{}
\label{sfig4:corrS5d_10ss}
\end{subfigure}
\begin{subfigure}[b]{0.49\textwidth}
\centering
\includegraphics[width=1\textwidth,keepaspectratio]{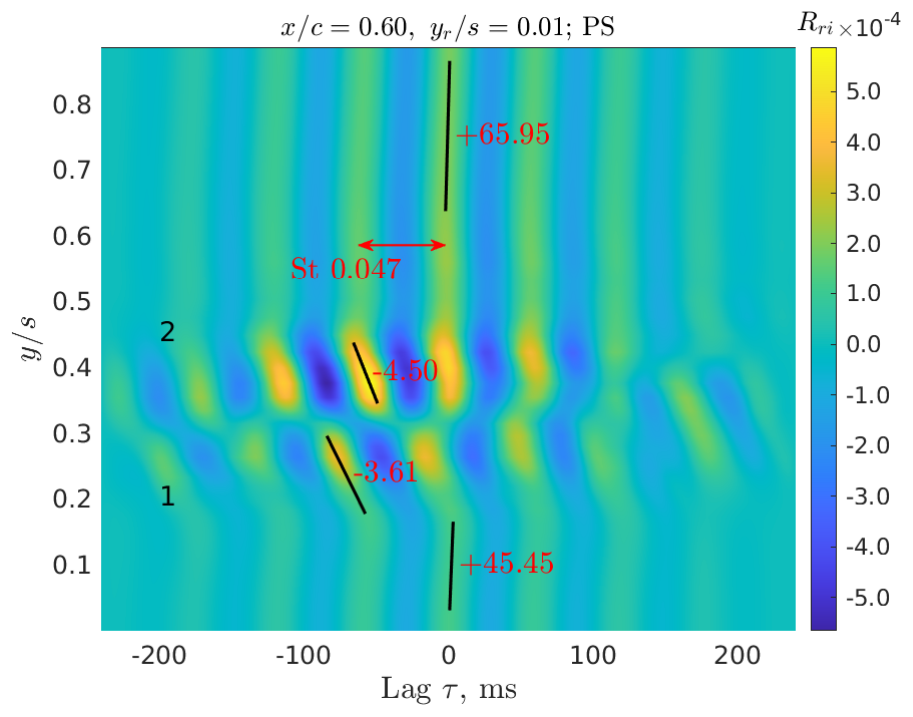}
\caption{}
\label{sfig4:corrS5d_10ps}
\end{subfigure}
\begin{subfigure}{0.49\textwidth}
\centering
\includegraphics[width=\linewidth]{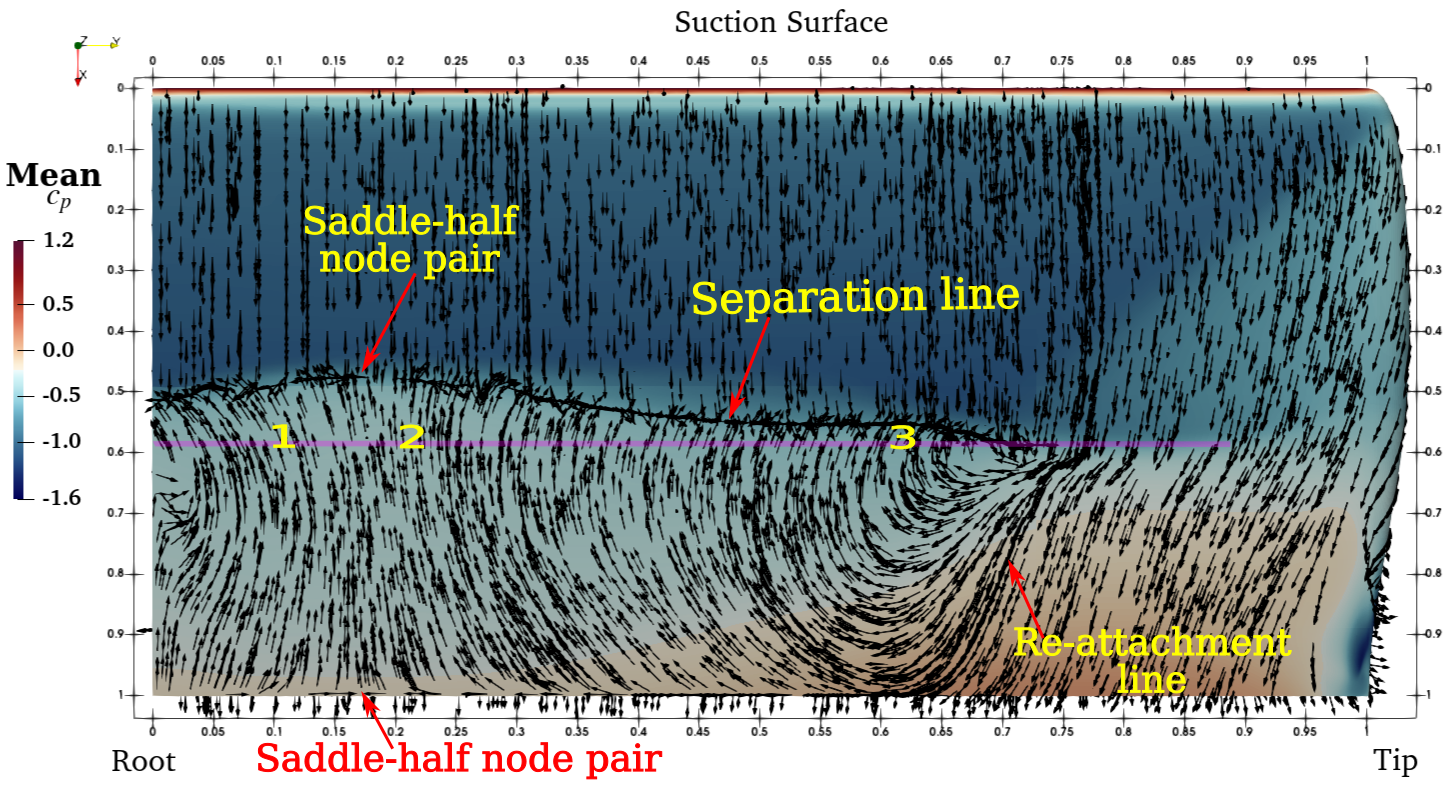}
\caption{}
\label{sfig4:msflns5dss}
\end{subfigure}
\begin{subfigure}{0.49\textwidth}
\centering
\includegraphics[width=\linewidth]{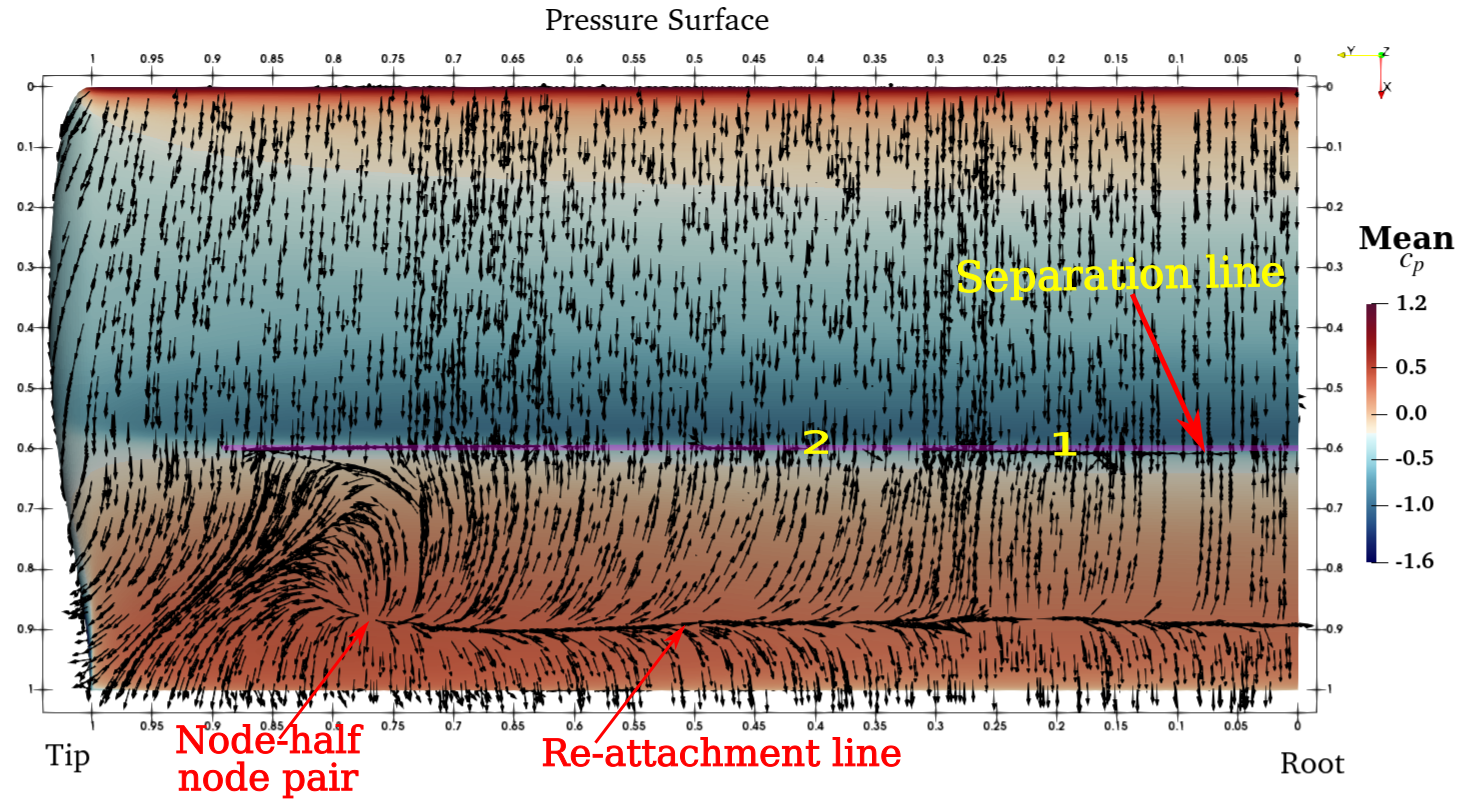}
\caption{} 
\label{sfig4:msflns5dps}
\end{subfigure}
\caption{$5^\circ$ AoA: Spanwise cross-correlations $R_{ri}$ (a,b) and mean skin-friction lines (c,d) on the suction and pressure surfaces}
\label{fig4:corrSpanw5d_10}
\end{figure}

\subsubsection{$5^\circ$}
\label{subsec:ch4_5d}

\begin{figure}[!ht]
\centering
\begin{subfigure}[b]{0.49\textwidth}
\centering
\includegraphics[width=1\textwidth,keepaspectratio]{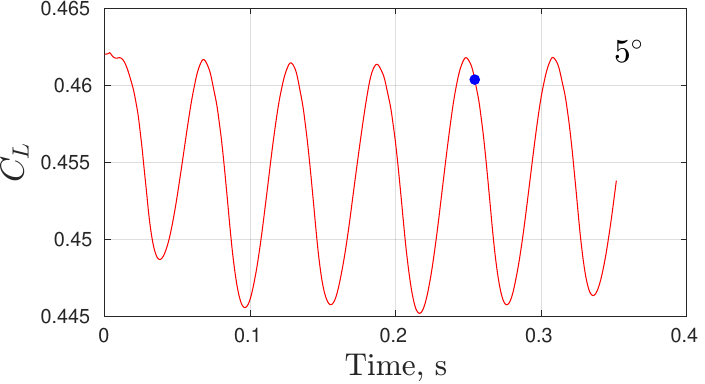}
\caption{}
\label{sfig4:ptOnCLSS5dlrgWL}
\end{subfigure}
\begin{subfigure}[b]{0.49\textwidth}
\centering
\includegraphics[width=1\textwidth,keepaspectratio]{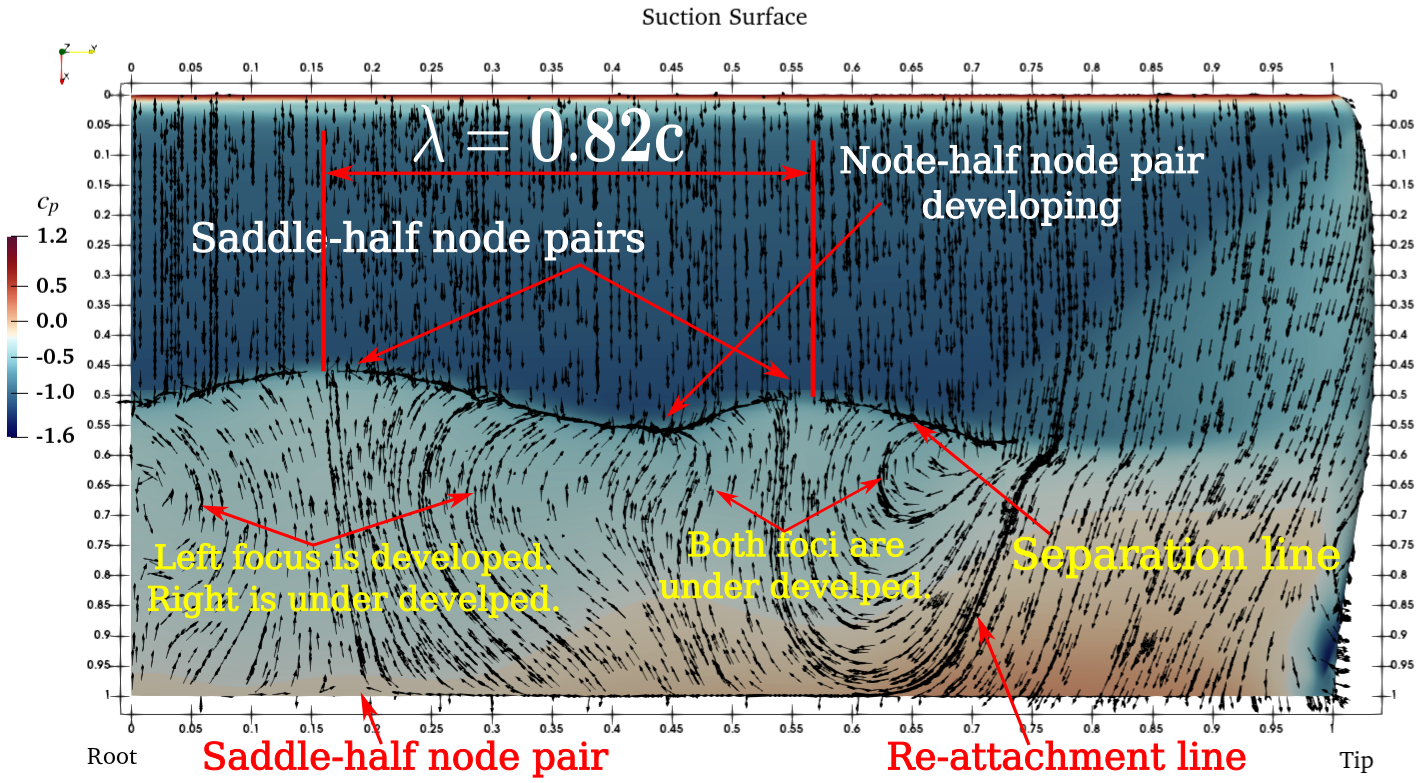}
\caption{}
\label{sfig4:waveLSS5dlrgWL}
\end{subfigure}
\caption{$5^\circ$ AoA: An instance on $C_L$ (a) and wavelength of propagating buffet cells at the same instance (b)}
\label{fig4:ptCLwaveLSS5dlrgWL}
\end{figure}

\begin{figure}[!ht]
\centering
\begin{subfigure}[b]{0.49\textwidth}
\centering
\includegraphics[width=1\textwidth,keepaspectratio]{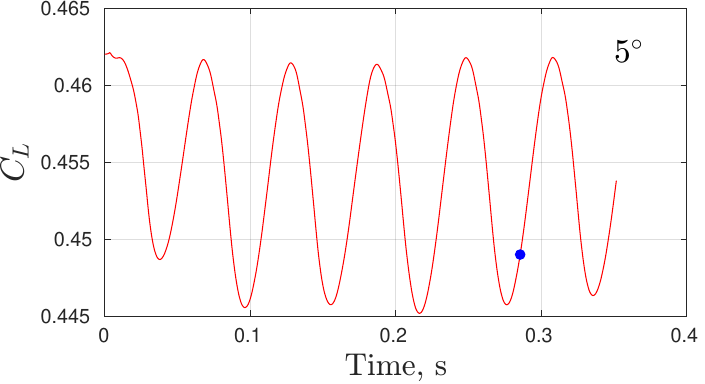}
\caption{}
\label{sfig4:ptOnCLSS5dsmlWL}
\end{subfigure}
\begin{subfigure}[b]{0.49\textwidth}
\centering
\includegraphics[width=1\textwidth,keepaspectratio]{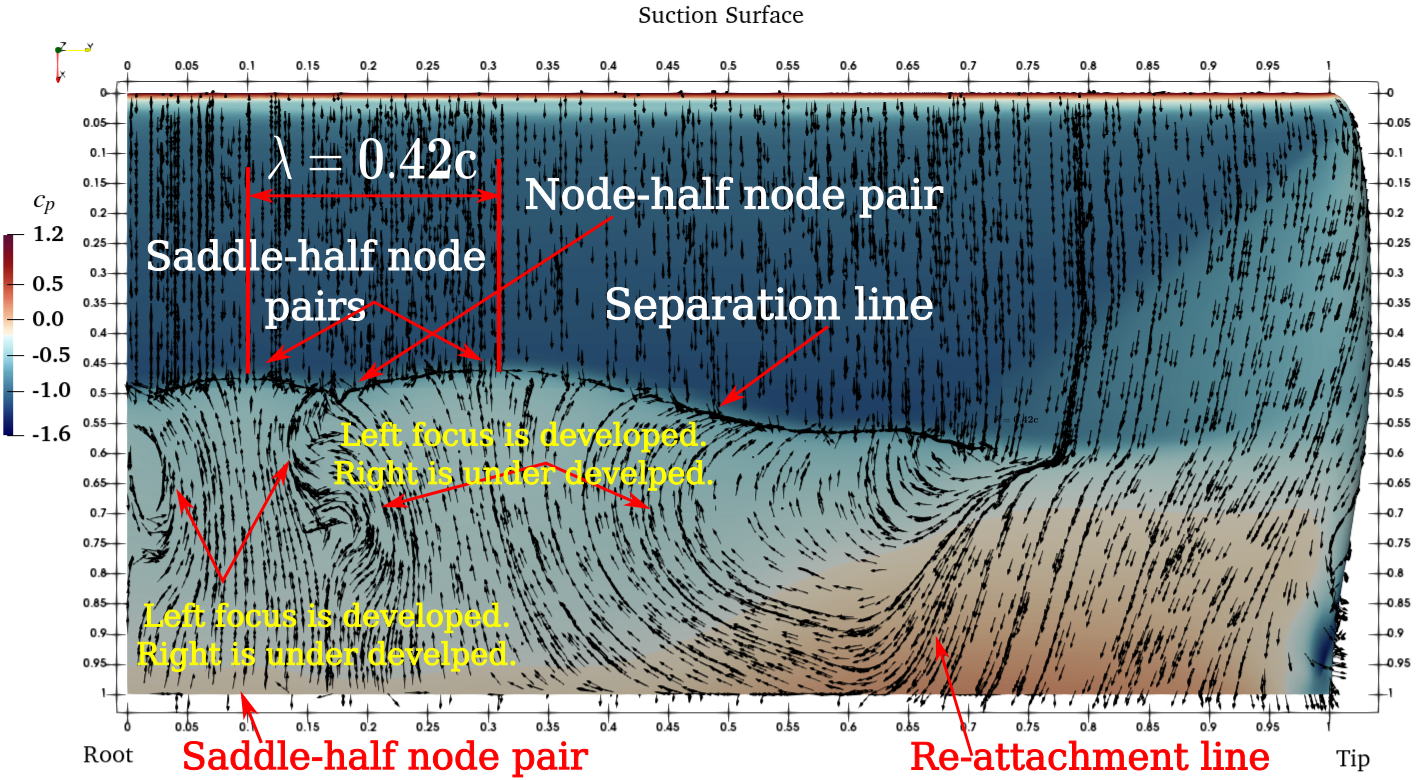}
\caption{}
\label{sfig4:waveLSS5dsmlWL}
\end{subfigure}
\caption{$5^\circ$ AoA: An instance on $C_L$ (a) and wavelength of propagating buffet cells at the same instance (b)}
\label{fig4:ptCLwaveLSS5dsmlWL}
\end{figure}

Spanwise correlations and mean skin-friction lines are given in \autoref{fig4:corrSpanw5d_10}. The correlation on SS is of order $\mathcal{O}(1)$ higher than that on PS. This relates to the proper propagation of inboard buffet cells along with self-induced motion of foci on SS, contrary to improper on PS. Wave propagation on SS is in a similar fashion as for $3^\circ$, as seen in Supplementary Video 5A. Here also, point $3$ location is an origin of both types of waves, traveling in opposite directions. Inboard hydrodynamic wave travels at velocity $-5.6$ m/s from $3$ to $2$ and acoustic at $16.15$ m/s outboard of $3$. Another hydrodynamic wave occurs from $2$ to $1$ at velocity $-2.9$ m/s with the same frequency $St~0.046$. Wavelength $\lambda_{pwp}=0.852c$ associated with $-5.60$ m/s agrees well with measured one $\lambda_{bp}=0.82c$ on skin-friction lines in \autoref{fig4:ptCLwaveLSS5dlrgWL}. And $\lambda_{pwp}=0.441c$ with $-2.90$ m/s agrees well with $\lambda_{bp}=0.42c$ on skin-friction lines in \autoref{fig4:ptCLwaveLSS5dsmlWL} at another instance. These topologies of critical points in these instances are underdeveloped. On PS, correlation is of higher magnitude and improper propagation only in between locations $1$ and $2$. Notice that partially developed foci and buffet cells in synchronization, travel inboard only between these locations in Supplementary Video 5B.

\section{Result's discussion}
\label{sec:discuss}
From time domain analysis, buffet offset occurs between $5^\circ$ and $7^\circ$ AoAs. Buffet-onset might occur below the $-1^\circ$ AoA, not explored in the present study. Buffet is shown to be a pre-stall aerodynamic instability. With increase of AoA, buffet amplitude and separation region typically increase on SS and decrease on PS. Increase or decrease of buffet amplitude relates to total higher or lower vorticity (sum of vorticities' magnitudes) of unstable contra-rotating foci in a pair, respectively. This is noticeable in the respective supplementary videos. Both contra-rotating foci of higher total vorticity magnitude drive the reverse flow at a higher speed. This creates a higher curvature on the separation line in contrast to the foci of lower total vorticity magnitude. For example, buffet amplitude on SS at $10\%$ span in \autoref{subfig:baVSaoa} increases from $0^\circ$ to $1^\circ$ as a pair of contra-rotating foci develops during this sweep of AoA in the same region. And then, it decreases from $1^\circ$ to $5^\circ$ as the pairs of foci lose vortex strength while reaching near-root from mid-span region for $3^\circ$ and $5^\circ$ AoAs. Notice also that they reduce in size during this inboard motion for these two AoAs.

From frequency domain analysis, 3D buffet Strouhal numbers $St$ are computed in order of $\mathcal{O}(-2)$. In literature, 2D buffet frequencies are found in range of $St$ $0.06-0.08$ ($\mathcal{O}(-2)$) \cite{deck2005,jacquin2009,hartmann2012time,arif2023two}. As seen here, the shock buffet frequencies are of the same order as 2D buffet frequencies. In this connection, \citet[Fig. 16 and 17]{dandois2016experimental} captured such low-frequencies of $St$ $0.02-0.08$ of inboard propagating disturbance on a swept wing. But, this was not conclusively established in this study. Another experimental study \citet{masini2020analysis} reported inboard propagating buffet cells associated with low frequencies of $St$ $0.05-0.15$ at pre and post-buffet conditions.

Analyzing the correlation results with $c_p$ distribution and skin-friction lines provides an intercorrelation among these plots. Marking on the correlation and $c_p$ distribution plots are special locations in the topology of critical points in skin-friction lines, as already shown and discussed. Streamwise maximum and minimum correlation values at regular intervals in the shock region indicate periodic oscillations of the shock for all AoAs except for $3^\circ$. Spanwise correlations typically exhibit inboard propagation of waves of a hydrodynamic nature. These waves are associated with the self-induced inboard motion of unstable foci. It is noticeable that acoustic waves propagate at higher speeds than the hydrodynamic waves in both, streamwise and spanwise, correlation plots. Streamwise propagation is found mostly in acoustic nature. Both hydrodynamic and acoustic spanwise propagations travel simultaneously in opposite directions at locations of mark $3$ on SS for $3^\circ$ and $5^\circ$ AoAs. Flow at $3^\circ$ AoA is predicted with special characteristics, such as highly unsteady flow, a band of frequencies, and inboard propagation of buffet cells on both surfaces, unlike other AoAs.

\begin{table}[h!]
\centering
\caption{Characteristic parameters of spanwise propagation} \label{tab:ISProp}
\begin{tabular}{c c c c c c}\\
\hline
AoA & Surface & $St$ & $V_p/U_{\infty}$ & $\lambda_{pwp}/c$ & $\lambda_{bp}/c$  \\[0.0ex]
\hline
\multirow{2}{2em}{$-1^\circ$} & SS & $0.085$~ & $-0.420$ & $4.9$ & no BCs \\[0.0ex]
 & PS & $0.086$~ & $-0.058$ & $0.67$ & $0.64$ \\[0.0ex]
\hline
\multirow{2}{2em}{$0^\circ$} & SS & $0.068$~ & $1.045$ & $15.35$ & no BCs \\[0.0ex]
 & PS & $0.068$~ & $-0.0532$ & $0.78$ & $0.74$ \\[0.0ex]
\hline
\multirow{3}{2em}{$1^\circ$} & SS & $0.076$~ & $0.187$ & $2.46$ & no PoBCs \\[0.0ex]
 & mid-span PS & $0.076$~ & $-0.0588$ & $0.77$ & $0.72$ \\[0.0ex]
  & near-root PS  & $0.076$~ & $-0.024$ & $0.32$ & $0.34$ \\[0.0ex]
\hline
\multirow{4}{2em}{$3^\circ$} & near-tip SS & $0.041$~ & $0.180$ & $4.40$ & no BCs \\[0.0ex]
 & mid-span SS & $0.041$~ & $-0.0273$ & $0.66$ & $0.62$ \\[0.0ex]
 & near-root SS & $0.041$~ & $-0.0163$ & $0.38$ & $0.39$ \\[0.0ex]
  & PS & $0.080$~ & $-0.0434$ & $0.542$ & $0.54$ \\[0.0ex]
\hline
\multirow{4}{2em}{$5^\circ$} & near-tip SS & $0.046$~ & $0.113$ & $2.46$ & no BCs \\[0.0ex]
 & mid-span SS & $0.046$~ & $-0.0392$ & $0.85$ & $0.82$ \\[0.0ex]
 & near-root SS & $0.046$~ & $-0.0203$ & $0.44$ & $0.42$ \\[0.0ex]
  & PS & $0.047$~ & not regular PoBCs & not regular PoBCs & not regular PoBCs \\[0.0ex]
\hline
\end{tabular}\\[0.8ex]
BCs: Buffet Cells;~~ PoBCs: Propagation of Buffet Cells
\end{table}

Characteristic parameters of spanwise propagation waves are summarized in \autoref{tab:ISProp} for all AoAs. All these parameters are non-dimensional. Notice that propagation velocity $V_p/U_{\infty}$ of order $\mathcal{O}(-2)$ is of hydrodynamic nature. They are associated with the propagation of buffet cells and fundamentally with the self-induced motion of unstable foci. That is why their wavelengths, $\lambda_{pwp}/c=(V_p/U_{\infty})/St$, match with the wavelengths $\lambda_{bp}/c$ of propagation of buffet cells.

\begin{figure}[!ht]
\centering
\begin{subfigure}[b]{0.49\textwidth}
\centering
\includegraphics[width=1\textwidth,keepaspectratio]{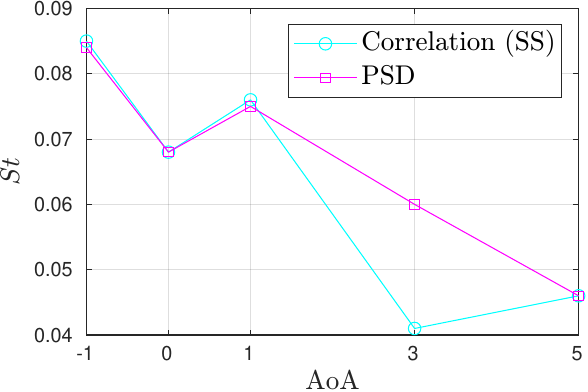}
\caption{}
\label{sfig:WPFvsAoAss}
\end{subfigure}
\hfill
\begin{subfigure}[b]{0.49\textwidth}
\centering
\includegraphics[width=1\textwidth,keepaspectratio]{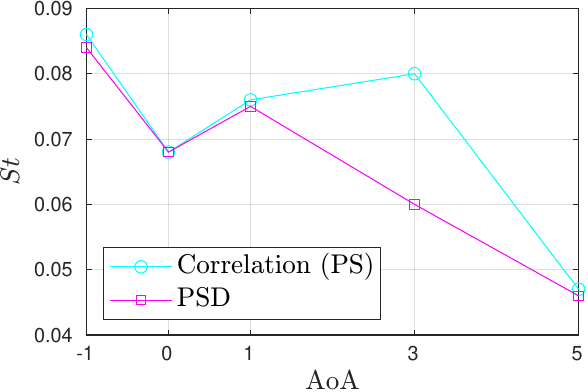}
\caption{}
\label{sfig:WPFvsAoAps}
\end{subfigure}

\begin{subfigure}[b]{0.49\textwidth}
\centering
\includegraphics[width=1\textwidth,keepaspectratio]{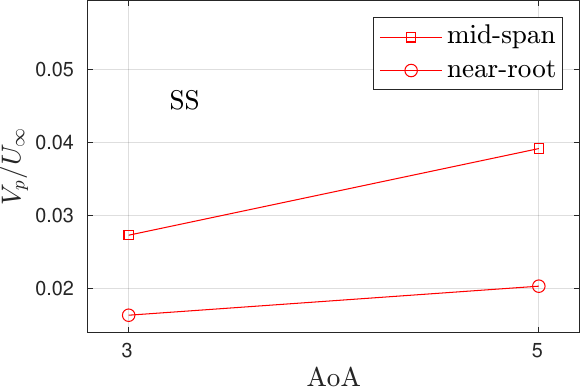}
\caption{}
\label{sfig:WPSvsAoAss}
\end{subfigure}
\hfill
\begin{subfigure}[b]{0.49\textwidth}
\centering
\includegraphics[width=1\textwidth,keepaspectratio]{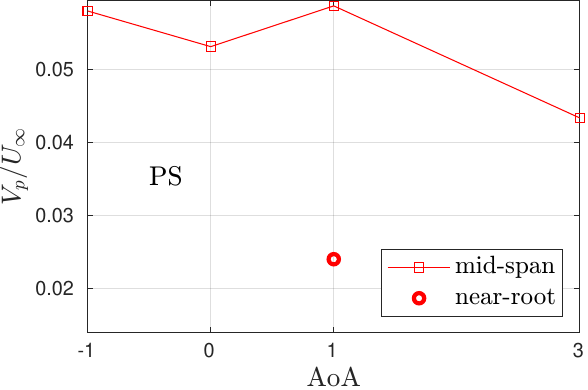}
\caption{}
\label{sfig:WPSvsAoAps}
\end{subfigure}

\begin{subfigure}[b]{0.49\textwidth}
\centering
\includegraphics[width=1\textwidth,keepaspectratio]{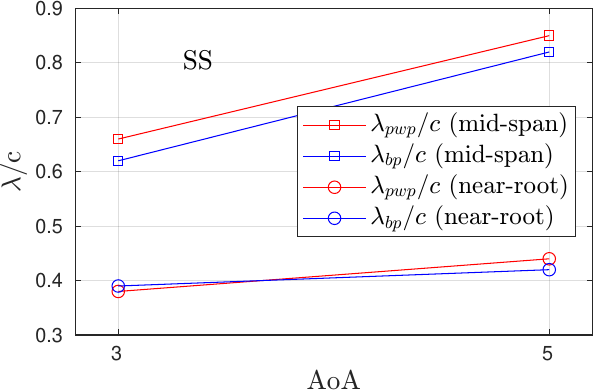}
\caption{}
\label{sfig:WLvsAoAss}
\end{subfigure}
\hfill
\begin{subfigure}[b]{0.49\textwidth}
\centering
\includegraphics[width=1\textwidth,keepaspectratio]{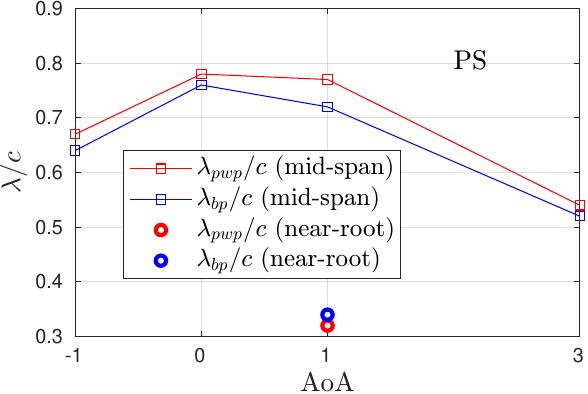}
\caption{}
\label{sfig:WLvsAoAps}
\end{subfigure}
\caption{Characteristic parameters variation with increase of AoA}
\label{fig:FrWPSwaveL}
\end{figure}

The characteristic parameters $St$, $V_p/U_{\infty}$, and $\lambda/c$ are plotted with respect to AoA in \autoref{fig:FrWPSwaveL}. Strouhal numbers from correlations and PSDs on both surfaces compare well to each other except $3^\circ$ AoA in \autoref{sfig:WPFvsAoAss} and \autoref{sfig:WPFvsAoAps}. Mismatch at $3^\circ$ is due to broadband spectrum $St~0.04-0.08$. In correlation plots, lower limit of this band is captured on SS, and higher limit is on PS. However, correlations and PSDs both show a typical decrease in frequency with respect to AoA. Inboard propagation velocities $V_p/U_{\infty}$ vary in the mid-span and near-root regions in \autoref{sfig:WPSvsAoAss} on SS and in \autoref{sfig:WPSvsAoAps} on PS. These waves are of a hydrodynamic nature on SS for $3^\circ$ and $5^\circ$ AoAs and on PS for $-1^\circ$, $0^\circ$, $1^\circ$, and $3^\circ$ AoAs. Propagation speeds in mid-span region on both surfaces are higher than those in the near-root region. On SS, propagation speed increases with AoA in both regions; on PS, it typically decreases in mid-span region. Notice that the speed in mid-span at $3^\circ$ is higher on PS than that on SS. At last, pressure wave propagation wavelengths $\lambda_{pwp}$ compare well with buffet cells propagation wavelengths $\lambda_{bp}$ in \autoref{sfig:WLvsAoAss} and \autoref{sfig:WLvsAoAps}. This match is obvious, as already mentioned, these pressure waves are of hydrodynamic nature, so carried out by 3D flow structures. The same is clearly seen in the respective supplementary videos. On SS, wavelength increases from $3^\circ$ to $5^\circ$ AoA in mid-span and near-root regions; on PS, it is almost constant and maximum at $0^\circ$ and $1^\circ$ AoAs than others in mid-span region. Wavelengths in near-root are smaller than that in the mid-span on both surfaces. Wavelengths relate to the size of buffet cells and fundamentally to the size of contra-rotating foci in pairs. Both, buffet cells and foci, reduce in size while reaching the root region from the mid-span and so the wavelengths on SS for $3^\circ$ and $5^\circ$ AoAs. The same is noticed in Supplementary Video 4A and 5A for $3^\circ$ and $5^\circ$ AoAs, respectively. For only $1^\circ$ AoA, on PS, small and large wavelengths are associated with small and large foci, respectively. They are present simultaneously on PS as seen in Supplementary Video 3B.

\subsection{Shock buffet mechanism}
\label{ssec:shockBuffMech}
\begin{figure}[!ht]
\centering
\includegraphics[width=\linewidth]{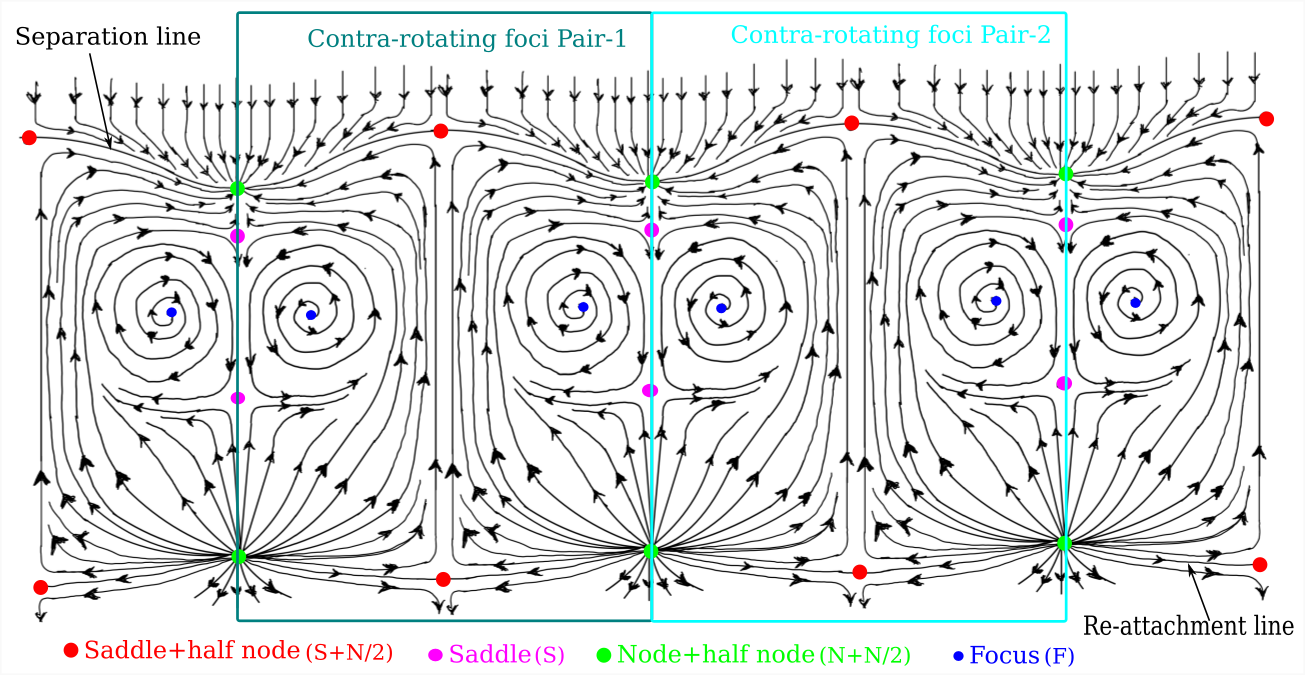}
\caption{A general flow topology in separated region}  
\label{fig:GFT}
\end{figure}

We have observed different flow topologies on both surfaces at different AoAs. These topologies refer to an interrelation of critical points on skin friction lines. Observing these topologies on both surfaces in Supplementary Videos 1 to 5 for each AoA, we propose a general topology shown in \autoref{fig:GFT}. This topology will change with AoA, Mach number, and Reynolds number, but always satisfies a relation among the critical points as given by \autoref{eq:saddle_nodesfoci_rel}.

\autoref{fig:GFT} shows separation and re-attachment lines enclosing the separated region. For every AoA, these lines are present on both surfaces. Separation line is a locus of flow separation positions along the span, which is almost the locus of shock foot along the span, as seen in the supplementary videos. re-attachment line is the locus of flow re-attachment position along the span. Separation line passes through critical points "Saddle$+$half-node (S+N/2)" on the crest and "Node$+$half-node (N+N/2)" on the trough. This wavy nature of separation line represents the formation of buffet cells. Separation line is also in form of a straight line, such as on the SS for $-1$ and $0^\circ$ AoAs. Re-attachment line is upstream or downstream of the trailing edge, passing through "Saddle$+$half-node (S+N/2)" and "Node$+$half-node (N+N/2)" critical points as following the separation line; otherwise, an inclined line on SS for $-1$, $0^\circ$, and $1^\circ$ AoAs. "Saddle$+$half-node" represents a coincidence of a saddle on the surface and a half-node in the flow above the surface \cite[Section 2.3]{delery2013}. Similarly, "Node$+$half-node" represents a coincidence of a node on the surface and a half-node in the flow. Within the region enclosed by the separation and re-attachment lines, unstable focus $F$, and saddle $S$ points are present.

\begin{figure}[!ht]
\centering
\begin{subfigure}[b]{0.49\textwidth}
\centering
\includegraphics[width=1\textwidth,keepaspectratio]{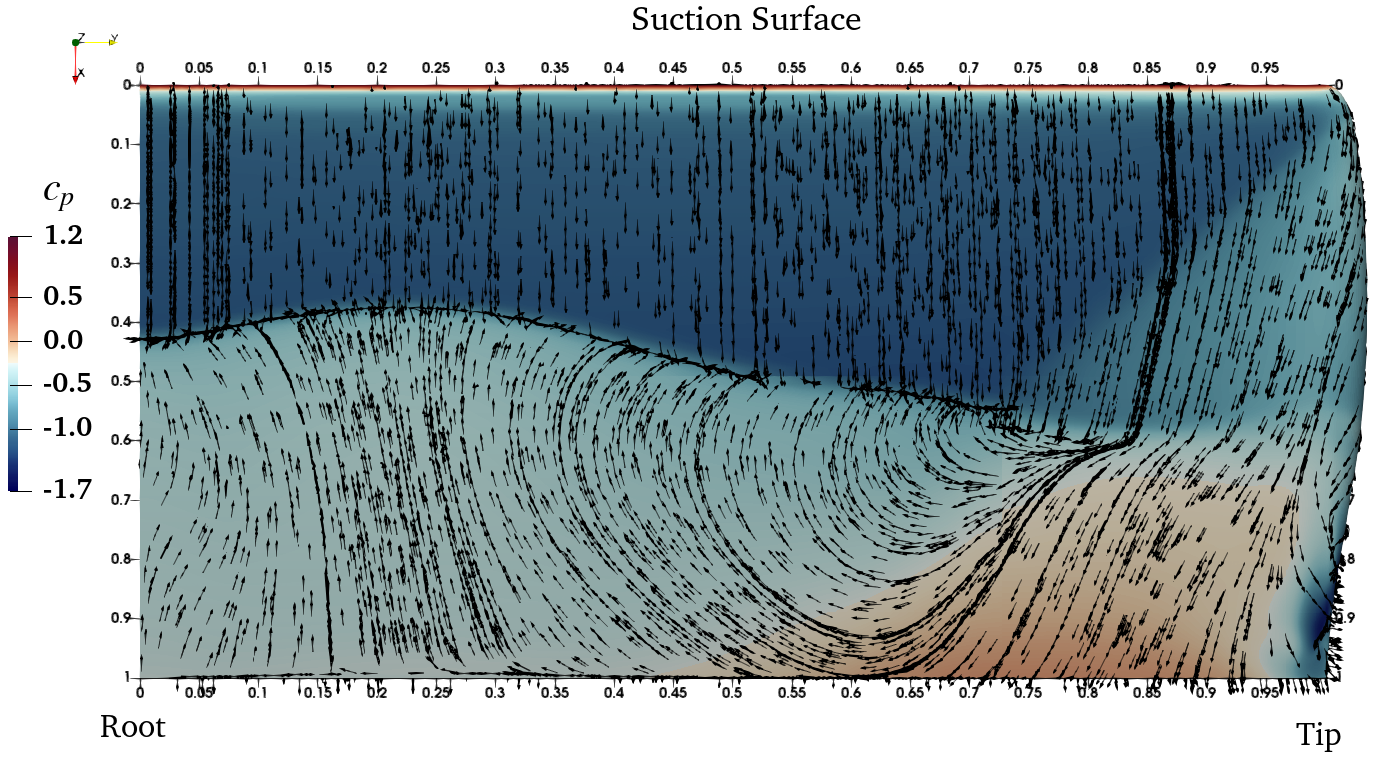}
\caption{}
\label{sfig:SFLn7degSS}
\end{subfigure}
\hfill
\begin{subfigure}[b]{0.49\textwidth}
\centering
\includegraphics[width=1\textwidth,keepaspectratio]{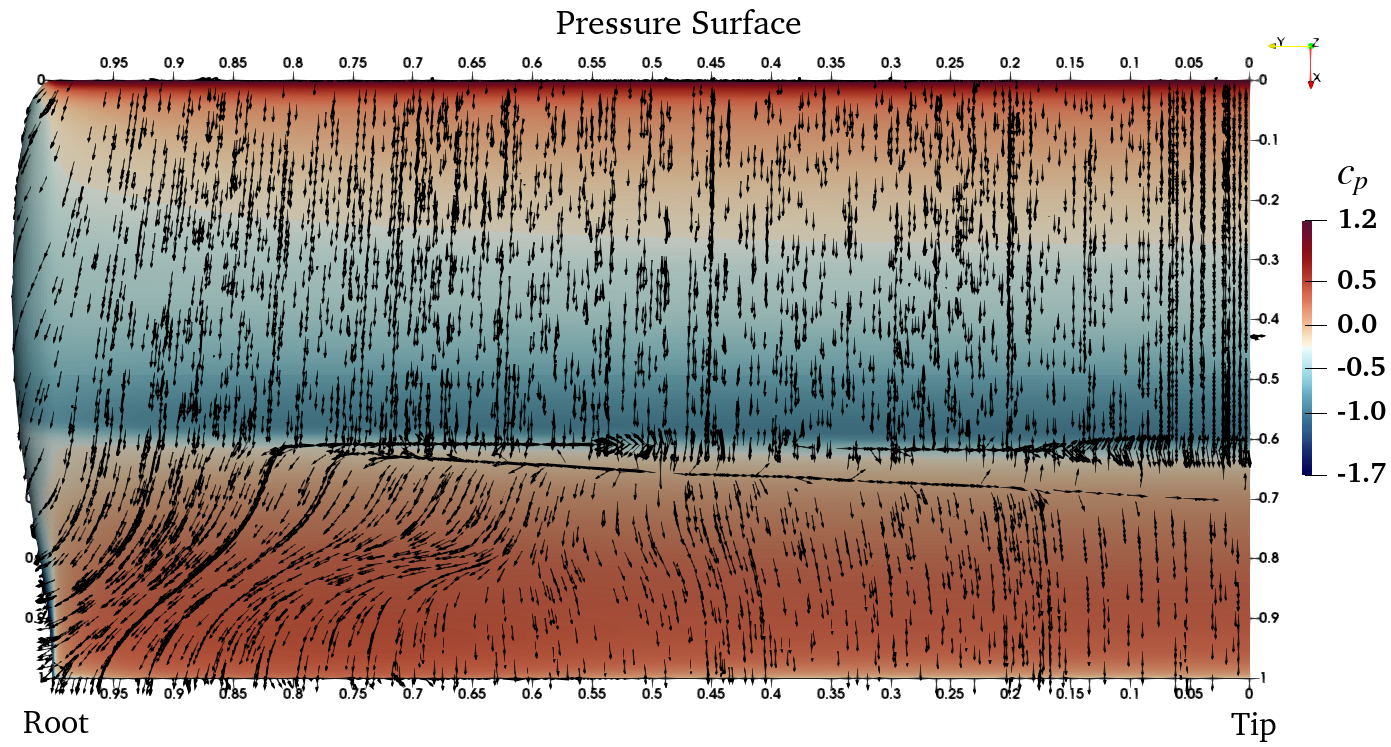}
\caption{}
\label{sfig:SFLns7degPS}
\end{subfigure}
\caption{Skin-friction lines over $c_p$ contours for $7^\circ$ AoA}
\label{fig:SFLns7deg}
\end{figure}

Two contra-rotating unstable foci are always associated with a curvature/buffet cell on the separation line. Such pairs (pair-1 and pair-2 boxes) are shown in general topology \autoref{fig:GFT}. They provide a clear picture of how two unstable conta-rotating foci drive the reverse flow against the separation line and create a curvature on the separation line. Such a pair has already been observed in an experimental study by \citet[Fig. 5]{jacquin2009}, but they concluded, "... they originate from stiff 2-D pressure waves produced by a more 3-D hydrodynamic organization in the separated flow." Such pairs are also indicated on infinite straight and swept wings in \citet[Figures 6,7, and 8]{plante2020similarities}. Particularly, notice pairs of stable and unstable contra-rotating foci for $10^\circ$ sweep in Figure 7. If intensity of skin-friction lines is increased in these figures, such pairs would be seen very clearly. Apart from this, \citet[Figure 17]{plante2020similarities} shows pairs of stable contra-rotating foci in flow topology of stall cells in subsonic regime. But they did not point out in their study that such structures are nothing but the critical points. On the contrary, such pairs are not seen on skin-friction lines in \autoref{fig:SFLns7deg} because of buffet disappearance at $7^\circ$ AoA or vice versa. 

General topology in \autoref{fig:GFT} is a fully developed state of the critical points, which satisfies the relation in \autoref{eq:saddle_nodesfoci_rel}. Within the pair-1 box, two foci and two saddles satisfy the relation. Combining pair-1 and pair-2 boxes, a total of 2 nodes, 4 foci, and 6 saddles again satisfies the relation. The same relation is satisfied for every instance of a fully developed topology of skin-friction lines in the supplementary videos. Such a fully-developed state in \autoref{sfig4:waveLPS0d} on PS for $0^\circ$ AoA satisfies the same relation as earlier mentioned in \autoref{subsec:ch4_0d}.

The separation line turns from a straight line to a fully wavy line on SS and from a fully wavy to a somewhat straight line on PS as AoA increases from $-1^\circ$ to $5^\circ$. This is clearly seen on the respective surface in Supplementary Videos 1 to 5. In case of wavy separation lines, the fully grown pairs of unstable foci move spanwise and generate a traveling wave on the separation line as shown in \autoref{fig:GFT}. These unstable foci travel because of their self-induced motion. The re-attachment line typically follows the separation line on both surfaces. It turns from a wavy line to a somewhat straight line on PS as AoA increases from $-1^\circ$ to $5^\circ$. On SS, it is an inclined line for $-1^\circ$, $0^\circ$, and $1^\circ$ AoAs and lies downstream of TE for $3^\circ$ and $5^\circ$ AoAs. Notice the same in the respective Supplementary Videos 1 to 5. Both lines, separation and re-attachment, initiate from a common point on both surfaces.

Critical point theory explains that an unstable focus accelerates the flow from its inner core to the outer region, contrary to a stable focus. Its unstable nature creates a self-induced motion on the surface. In the present investigation, this inboard motion of contra-rotating unstable foci in pairs causes inboard propagation of buffet cells on the separation line on both surfaces. Particularly, inboard motion of these foci might be the result of spanwise favorable pressure gradient in the separated region from tip to the root, observed in the respective supplementary videos. Their motion and interaction with each other need to be investigated thoroughly. However, change in pressure distribution on the surface is apparent with the movement of the critical points in a topology in supplementary videos. For example, pressure is always lower in foci region than in node$+$half-node (N+N/2) and saddle$+$half-node (S+N/2) points on re-attachment line during their motion. Refer to the respective Supplementary Videos 1B, 2B, 3B, and 4B for observing the same phenomenon. The same propagation of pressure synchronized with the motion of critical points is captured in spanwise correlation plots.

\section{Conclusions}
\label{sec:conclu}

The buffet frequencies for an unswept finite-span benchmark supercritical wing in flow at Mach $0.85$ and Reynolds number $4.491\times10^6$ are found comparable to 2D buffet frequencies. In the present study, buffet frequency typically decreases with AoA, unlike increase in 2D buffet. However, buffet amplitude is much lower relative to that of the 2D buffet, as mentioned in the literature. The shock amplitude on the suction surface increases with AoA, but decreases on the pressure surface. $3^\circ$ is a special AoA with a band of frequencies and propagation of buffet cells on both surfaces, unlike $-1^\circ$, $0^\circ$, $1^\circ$, and $5^\circ$ AoAs. For these AoAs, 3D structures in separation regions are nothing but critical points---nodes, unstable foci, and saddles---in a certain topology. The motion of these critical points changes the pressure distribution on the surface which reflects as hydrodynamic wave propagation in correlation plots. Pairs of contra-rotating unstable foci are found in the topology of skin-friction lines, responsible for the propagation of buffet cells. One or multiple such pairs create one or multiple buffet cells on the separation line. The self-induced motion of these unstable foci in pairs generates the propagation of buffet cells. Such contra-rotating foci have already been shown in numerical and experimental buffet studies, but it was not pointed out that they are the critical points contributing to the propagation of buffet cells. These pairs of foci influence the propagation of buffet cells in two aspects: first, total vorticity magnitude of both contra-rotating foci influences the buffet amplitude; second, sizes of these contra-rotating foci influence the wavelength of propagating buffet cells. When self-induced motion of unstable foci in pairs is absent or confined to a particular region, then buffet cells are also not present or confined to that corresponding region such as buffet cells on the suction surface for $-1^\circ$, $0^\circ$, and $1^\circ$ AoAs, and on the pressure surface for $5^\circ$ AoA.

The narratives on transonic buffet and its cause: be it a feedback mechanism of downstream and upstream propagating pressure waves; modal analysis of transonic buffet onset revealing that the mode characterizing shock and boundary layer interaction whose damping has just turned negative; global stability analysis revealing that onset of shock oscillations are governed by a linear unstable mode and limited by nonlinear saturation, all have their place and importance in the pantheon of perspectives on transonic buffet. The perspective presented here, based on the unsteady evolution of the critical points of the skin friction lines over suction and pressure surface of the wing causing shock oscillations, can be the basis of all the mechanisms presented earlier in the literature on transonic buffet. For one, for a wing of a given span, sweep, and airfoil section at a given Mach and Reynolds number, and angle of attack, the nature and distribution of critical points are distinct. These critical points define the nonlinear fluid dynamics causing the transonic shock oscillations. It helps to remember that one is dealing with a very complex nonlinear dynamical system; the flow parameters as well as the wing planform and section geometry, all play very significant roles in characterizing the rich and varied aerodynamics of transonic buffet response of the wing.

\bibliography{mag}

\end{document}